\documentclass[12pt,a4paper]{article}
\usepackage[english]{babel}  
\usepackage[margin=1in]{geometry}
\RequirePackage[OT1]{fontenc}
\RequirePackage{graphicx}
\RequirePackage{amsthm,amssymb}
\RequirePackage[cmex10]{amsmath}
\RequirePackage[colorlinks,citecolor=blue,urlcolor=blue]{hyperref}

\usepackage[round,authoryear]{natbib}
\usepackage{graphicx,epstopdf}
\usepackage{listing}
\usepackage{amsfonts}
\usepackage{amsmath}		
\usepackage{bbm} 	
\usepackage{accents}
\usepackage{amsmath,bm}
\usepackage{mathtools}
\usepackage{booktabs}
\usepackage{xcolor}
\usepackage{tabu}\usepackage{amsmath}
\usepackage{amsfonts}
\usepackage{amssymb}

\usepackage{caption} 
\usepackage{verbatim}
\usepackage{multirow}
\usepackage{authblk}
\usepackage[multiple]{footmisc}
\usepackage{easy-todo}

\usepackage{color}
\usepackage{array}
\definecolor{dkgreen}{rgb}{0,0.6,0}
\definecolor{gray}{rgb}{0.5,0.5,0.5}
\definecolor{mauve}{rgb}{0.58,0,0.82}

\usepackage{xr}
\makeatletter
\newcommand*\rel@kern[1]{\kern#1\dimexpr\macc@kerna}
\newcommand*\widebar[1]{%
  \begingroup
  \def\mathaccent##1##2{%
    \rel@kern{0.8}%
    \overline{\rel@kern{-0.8}\macc@nucleus\rel@kern{0.2}}%
    \rel@kern{-0.2}%
  }%
  \macc@depth\@ne
  \let\math@bgroup\@empty \let\math@egroup\macc@set@skewchar
  \mathsurround\z@ \frozen@everymath{\mathgroup\macc@group\relax}%
  \macc@set@skewchar\relax
  \let\mathaccentV\macc@nested@a
  \macc@nested@a\relax111{#1}%
  \endgroup
}
\makeatother

\newtheorem{assumption}{Assumption}

\newtheorem{proposition}{Proposition}[section]
\newtheorem{corollary}{Corollary}[section]
\newtheorem{theorem}{Theorem}[section]

\theoremstyle{remark}
\newtheorem{remark}{Remark}[section]

\newcommand{\power}{\pi}
\newcommand{\test}{\phi}
\newcommand{\testT}{\phi^{(\n)}}
\newcommand{\fptb}{b}
\newcommand{\vfptb}{\boldsymbol{\fptb}}
\newcommand{\e}{\varepsilon}
\newcommand{\ve}{\boldsymbol{\e}}

\newcommand{\vv}{\mathbf{v}}
\newcommand{\ii}{t}
\newcommand{\n}{T}
\newcommand{\p}{p}

\newcommand{\vx}{\mathbf{x}}
\newcommand{\y}{y}
\newcommand{\vy}{\mathbf{y}}
\newcommand{\vyy}{\vy_{t-1,1}}
\newcommand{\f}{f}

\newcommand{\Sig}{\boldsymbol{\Sigma}}

\newcommand{\J}{\boldsymbol{J}}

\newcommand{\id}{\mathbf{I}}
\newcommand{\mfzero}{\boldsymbol{0}}
\newcommand{\mi}{\mathcal{M}}
\newcommand{\dct}{\mfd_{\mfC,\ii}}
\newcommand{\dcu}{\mfd_{\mfC}(\s)}
\newcommand{\dcub}{\widebar{\mfd}_{\mfC}}

\newcommand{\ploc}{\boldsymbol{\mu}}

\newcommand{\ptrd}{\boldsymbol{\tau}}
\newcommand{\lptrd}{\boldsymbol{\delta}}
\newcommand{\lpptb}{\boldsymbol{\eta}}

\newcommand{\LR}{\mathcal{L}}
\newcommand{\CS}{\mathit{\Delta}}
\newcommand{\QV}{\mathcal{Q}}
\newcommand{\E}{\mathrm{E}}      
   
\newcommand{\Elim}{\mathbb{E}}   
\newcommand{\R}{\mathbb{R}}
\newcommand{\C}{C}
\newcommand{\N}{\mathbb{N}}
\newcommand{\F}{\mathfrak{F}}
\newcommand{\dd}{{\rm d}} 
\newcommand{\var}{\operatorname{Var}}
\newcommand{\cov}{\operatorname{Cov}}

\newcommand{\pto}{\stackrel{p}{\to}}
\newcommand{\wto}{\Rightarrow}

\newcommand{\law}{\mathrm{P}^{(\n)}}
\newcommand{\prob}{\mathbb{P}}
\newcommand{\trace}{{\rm Tr}}
\newcommand{\trans}{\prime}

\newcommand{\indicator}{\mathbbm{1}}

\newcommand{\cv}{\kappa}

\newcommand{\bias}{\boldsymbol{\zeta}}
\newcommand{\score}{\boldsymbol{\ell}}
\newcommand{\s}{u}
\newcommand{\W}{\boldsymbol{W}}
\newcommand{\We}{\W_{\ve}}
\newcommand{\Wehat}{\widehat\W_{\ve}}

\newcommand{\Wlf}{\W_{\score_\f}}

\newcommand{\Wb}{\W_{\vfptb}}

\newcommand{\Z}{\boldsymbol{Z}}
\newcommand{\Ze}{\Z_{\ve}}

\newcommand{\Zlf}{\Z_{\score_\f}}

\newcommand{\Zb}{\Z_{\vfptb}}

\newcommand{\B}{\boldsymbol{B}}
\newcommand{\Be}{\B_{\ve}}

\newcommand{\Blf}{\B_{\score_\f}}
\newcommand{\Blfhat}{\widehat\B_{\score_\f}^{(\n)}}

\newcommand{\Bb}{\B_{\vfptb}}

\newcommand{\WTe}{\W^{(\n)}_{\ve}}

\newcommand{\experiment}{\mathcal{E}}
\newcommand{\filtration}{\mathcal{F}}
\newcommand{\borel}{\mathcal{B}}

\newcommand{\mfC}{\mathbf{C}}
\newcommand{\mfB}{\mathbf{B}}
\newcommand{\mfA}{\mathbf{A}}
\newcommand{\mfd}{\mathbf{d}}
\newcommand{\mfG}{\mathbf{G}}
\newcommand{\mfD}{\mathbf{D}}

\newcommand{\mPi}{\boldsymbol{\Pi}}
\newcommand{\mGamma}{\boldsymbol{\Gamma}}
\newcommand{\malpha}{\boldsymbol{\alpha}}
\newcommand{\mbeta}{\boldsymbol{\beta}}

\newcommand{\mkg}{\mathfrak{g}}
\newcommand{\mkG}{\mathfrak{G}}

\newcommand{\mfc}{\mathbf{c}}

\graphicspath {{figures/}}

\title{Semiparametrically Optimal Cointegration Test}
\author{Bo Zhou}
\affil{Department of Economics, Virginia Tech}
\date{}

\begin{document}

\maketitle

\abstract{This paper aims to address the issue of semiparametric efficiency for cointegration rank testing in finite-order vector autoregressive models, where the innovation distribution is considered an infinite-dimensional nuisance parameter. Our asymptotic analysis relies on Le Cam's theory of limit experiment, which in this context takes the form of Locally Asymptotically Brownian Functional (LABF). By leveraging the structural version of LABF, an Ornstein-Uhlenbeck experiment, we develop the asymptotic power envelopes of asymptotically invariant tests for both cases with and without a time trend. We propose feasible tests based on a nonparametrically estimated density and demonstrate that their power can achieve the semiparametric power envelopes, making them semiparametrically optimal. We validate the theoretical results through large-sample simulations and illustrate satisfactory size control and excellent power performance of our tests under small samples. In both cases with and without time trend, we show that a remarkable amount of additional power can be obtained from non-Gaussian distributions.}



\section{Introduction} \label{sec:introduction}
Cointegration has been a central topic in time series econometrics ever since its concept was introduced by \citet{granger1981some} and \citet{engle1987co}. Cointegration refers to the phenomenon where multiple nonstationary time series have stationary linear combinations, known as cointegrating relationships. Determining the number of these relationships, or the cointegration rank, is of utmost importance. This inferential problem is often addressed in a finite-order vector autoregressive (VAR) process. Early attempts include, among others, residual-based tests by \citet{engle1987co} and \cite{phillips1990asymptotic}, likelihood ratio tests by \cite{johansen1988statistical,johansen1991estimation} and \cite{johansen1990maximum}, and principal component tests by \cite{stock1988testing}. Since then, extensive literature has been devoted to constructing cointegration rank tests with good power properties. Our paper shares the same goal, particularly investigating to what extent we can exploit testing power from the innovation distributions that deviate from Gaussianity.

The main objective of this paper is to develop semiparametrically optimal cointegration rank tests for a finite-order VAR model written in the ECM form. We assume that the innovations are independently and identically distributed, and we treat the innovation distribution as an infinite-dimensional nuisance parameter. Our analysis relying on Le Cam's asymptotic theory, where the concept of \textit{limit experiment} --- as the limit of the sequence of experiments of interest (here, the cointegration experiments) --- plays the central role; see, e.g., \cite{le2012asymptotic} and \cite{vdVaart00}. The most recent work using this approach is \cite{hallin2016semiparametric} (hereafter referred to as HvdAW), which provides a complete factorization for the cointegration parameter $\mPi$ in model given by (\ref{eqn:model_1})--(\ref{eqn:model_2}) below and characterizes all possible limit experiments. HvdAW shows that the associated limit experiments are of different types, including Locally Asymptotically Normality (LAN), Locally Asymptotically Mixed Normality (LAMN), and Locally Asymptotically Brownian Functional (LABF), depending on the parameter directions; see \cite{Jeganathan1995} for their definitions. HvdAW focuses on the LAN experiment brought by the time trend term, while some earlier works, such as \cite{phillips1991optimal} and \cite{hodgson1998adaptiveET,hodgson1998adaptiveJOE}, concentrate on on the LAMN direction. However, the LABF direction remains unexplored, and our aim is to fill this gap in the literature.

Our contribution to the literature is threefold. First, following \cite{zhou2019semiparametrically}'s structural representation technique for the LABF-type experiments, we develop the semiparametric power envelopes of asymptotically invariant tests in both cases with and without a time trend. Specifically, in this structural versions as an Ornstein-Uhlenbeck (OU) experiment, the nuisance density perturbation parameter ($\lpptb$) appears as a constant drift, which can be eliminated by taking the associate `bridge' process. More importantly, we show that the $\sigma$-field consisting of OU processes (for elements unaffected by $\lpptb$) and OU bridges (for elements affected by $\lpptb$) is maximally invariant, which provides optimality in the limit. According to the Neyman-Pearson lemma, the corresponding likelihood ratio test is optimal among all invariant tests since every invariant statistic is a function of the maximal invariant. The Asymptotic Representation Theorem then translates the limiting optimality to the sequence of cointegration experiments (see, e.g., \citet[Theorem 15.1]{vdVaart00}).

Our second contribution is to propose feasible tests whose powers can attain the semiparametric power envelopes asymptotically (as shown in Theorem~\ref{thm:liki_convergence} and Corollary~\ref{cor:optimaltests}). This confirms that the derived semiparametric power envelopes are indeed `envelopes', rather than just upper bounds, and that our tests are semiparametrically optimal. To construct these tests, we follow the traditional semiparametric inference literature, where the unknown nonparametric part is replaced by its kernel estimates (as see in works such as \citet{Bickel1982}, \citet{schick1986asymptotically}, and \cite{Klaassen1987}). To improve finite-sample performance, especially when the sample size is small, we adopt \cite{schick1987note}'s technique and use all samples for our semiparametric statistic without sample splitting. Our Monte Carlo study confirms the validity and optimality of our tests using large-sample simulations and exhibits satisfactory size control and excellent power performance in small-sample scenarios.

We regard the treatment of the linear time trend specification as our third contribution. Following the unit root and cointegration literature, we augment the stochastic term with a deterministic time trend term as in (\ref{eqn:model_1}). This additive structure, employed by many later works (e.g., \cite{elliott1996efficient} and \cite{Jansson2008} for unit root testing, and \cite{saikkonen2000trend}, \cite{lutkepohl2000testing}, and \cite{boswijk2015improved} for cointegration), has several advantages, including the ability to emphasize that the trend is at most linear. This trend specification is one of the main differences between our paper and HvdAW.\footnote{Another distinction of our paper from HvdAW is in terms of distribution assumption: HvdAW assumes the innovation distribution to be elliptical, while we allow it to be essentially unrestricted (up to the regularity conditions outlined in Assumption~\ref{assm:density_fp} for our asymptotic analysis). The reason for this difference is that HvdAW's proposed distribution-free rank-based tests utilize the concept of Mahalanobis distance, which in this context is for multidimensional generalization of rank-based inference under the more restricted elliptical distribution assumption.} HvdAW employs the traditional trend-in-VAR representation as in (\ref{eqn:model_h}), making the time trend the main power source for their test (essentially due to the super-consistency rate $T^{-3/2}$ introduced by the time trend). In contrast, our paper regards the time trend (local) parameter $\lptrd$ as a nuisance parameter to be eliminated (see Remark~\ref{remark:HvdAW} for more detailed discussions). We rely on the ``profile likelihood'' approach to eliminate this parameter, taking advantage of the simple quadratic structure of the likelihood with respect to $\ptrd$. Additionally, we develop a limiting statistic $\LR^{\ptrd*}_\f(\bar\mfC;\bar\mfC^*)$ for the semiparametric case, analogue to the $\mathit{\Lambda}_{p,C}^{GLS}(\bar{C};\bar{C}^*)$ statistic proposed by \citet{boswijk2015improved} for the Gaussian case. The statistic $\mathit{\Lambda}_{p,C}^{GLS}(\bar{C};\bar{C}^*)$ embeds and thus can help compare many existing Gaussian cointegration tests. Likewise, our semiparametric version $\LR^{\ptrd*}_\f(\bar\mfC;\bar\mfC^*)$ enables us to develop semiparametric versions of existing Gaussian tests that handle the time trend specification.

Finally, we discuss extensions that incorporate serially correlated errors and allow for more general reduced rank hypotheses, significantly expanding the applicability of our developed tests. Our discussion is based on existing limit experiment results, mostly provided by HvdAW, particularly their full limit experiment characterization in Proposition A.2 of their online supplementary appendix. Building on these results, we show that the inference for the parameter of interest, $\mPi$, is adaptive to both the parameters governing the serial correlation in the errors and those governing the existing cointegrating relationships. Therefore, we `just' need to consistently estimate these parameters and replace them with their estimates.

The remainder of the paper is structured as follows. Section~\ref{sec:model_setup} presents the model setup and assumptions. Section~\ref{sec:semipowerenvelope} develops the limit experiment,  sequentially eliminates the density perturbation and time trend parameters, and derives the corresponding semiparametric power envelopes for the cases with and without a time trend. Then, based on a nonparametrically estimated density, Section~\ref{sec:semi_inference} proposes feasible semiparametrically optimal tests, whose finite-sample performances are accessed by a Monte Carlo study in Section~\ref{sec:montecarlo}. Section~\ref{sec:extensions} provides discussions on necessarily extensions to expand the empirical applicability of our tests. The proofs of our theoretical results can be found in the supplementary Appendices.



\section{Model} \label{sec:model_setup}
We consider observations $\vy_1,\dots,\vy_{\n} \in \R^{\p}$ generated by the vector auto-regression (VAR) of order one in \textit{error correction form}
\begin{align} 
\vy_\ii =&~ \ploc + \ptrd\ii + \vx_\ii, \label{eqn:model_1} \\
\Delta\vx_\ii =&~ \mPi\vx_{\ii-1} + \ve_\ii, ~~~~~ \ii=1,\dots,T, \label{eqn:model_2}
\end{align}
where $\Delta$ denotes the first-order difference operator (i.e., $\Delta\vx_\ii = \vx_\ii - \vx_{\ii-1}$), $\ploc$ and $\ptrd$ (both in $\R^\p$) are unknown parameters that govern the constant and linear time trend terms, respectively, $\mPi\in\R^{\p\times\p}$ is an unknown parameter of interest, and $\{\ve_\ii\}$ is a $\p$-dimensional i.i.d.\ sequence of innovations with density $f$.

Throughout this paper, we assume $\vx_0 = \mfzero$ as the initial value condition. However, this assumption is less innocent than it may seem, as noted by \cite{elliott1996efficient} (ERS), who observed that even asymptotically, the initial observations can carry information. Further investigations into this issue can be found in \cite{muller2003tests} and \cite{elliott2006minimizing}. That being said, since our paper employs the same local-to-unity asymptotics as ERS, we can relax this condition to the same extent.

In this paper, we use the notation $\F_\p$ to represent the family of densities that satisfy the following assumptions, which we impose on $f$.

\smallskip
\begin{assumption}[]\label{assm:density_fp} 
\indent
\begin{enumerate} 
\item[(a)] $\f$ is absolutely continuous with a.e.\ gradient $\,\nabla \f(\ve_1) = \big(\partial\f(\ve_1)/\partial\e_{1,1},\dots,\partial\f(\ve_1)/\partial\e_{1,p}\big)^\trans$. 
\item[(b)] $\E_{f}[\ve_t] = \mfzero$ and the covariance $\Sig := \var_\f[\ve_1]$ is positive definite and finite. 
\item[(c)] The Fisher information
$\J_\f := \E_\f[\score_\f(\ve_1)\score_\f(\ve_1)^{\trans}]$, where $\score_\f(\ve_1) := -\nabla\f(\ve_1)/\f(\ve_1)$ denotes the location score of $f$, is finite.
\item[(d)] $\f$ is positive.
\end{enumerate}
\end{assumption}
\smallskip

The \textit{absolute continuity} assumption (a) on $f$ is a mild smoothness condition commonly imposed in the semiparametric literature. This condition is imposed for two reasons: Firstly, it allows us to proceed with the limit experiment approach (\citet{le2012asymptotic}) since it implies the \textit{differentiability in quadratic mean (DQM)} result, which is the exactly right condition needed for our log-likelihood ratio expansion (\citet{vdVaart00}).\footnote{A deep discussion and appreciation of DQM can be found in \cite{pollard1997another}.} Secondly, the absolute continuity assumption enables us to perform nonparametric estimation of the score function $\score_\f$, which will be used to construct a feasible semiparametrically optimal test. The finite-variance condition (b) guarantees that the Fisher information matrix $\J_\f$ is nonsingular (\citet[Theorem 2.3]{mayer1990cramer}). Together with the finite-Fisher-information condition (c), they ensure the the weak convergence of the partial-sum processes of $\ve_t$ and $\score_\f(\ve_t)$ to Brownian motions. The positive density condition (d) is merely for notational convenience, e.g., when defining the score function $\score_\f$.

As the primary focus of this paper is to address the issue of semiparametric efficiency in the context of cointegration, we begin by considering a simple case of testing the null hypothesis of no cointegration against the alternative of the existence of at least one cointegrating relationship, stated as:
\begin{align}
H_0: \, \mPi = \mfzero {\rm ~~~(equivalently}, {\rm ~rank}\,(\mPi) = 0).
\end{align}
In Section~\ref{sec:extensions}, we will briefly discuss the extension of our results to more general reduced rank hypotheses on $\mPi$, drawing upon existing literature.

\smallskip 
\begin{remark} \label{remark:HvdAW}
The choice of time trend specification can have significant implications for the asymptotic results of cointegration tests. In this paper, we adopt the specification from the branch of unit root and cointegration literature which adds a level constant and a linear time trend to the stochastic component, as given in (\ref{eqn:model_1}). See, for example, \cite{elliott1996efficient} for unit root testing, and \cite{lutkepohl2000testing} and \cite{boswijk2015improved} for cointegration testing. It differs from the specification used in \cite{hallin2016semiparametric}, which follows the traditional ``trend-in-VAR'' specification given by
\begin{align} \label{eqn:model_h}
\Delta\vy_\ii = \mathbf{v} + \mathbf{v}_1\ii + \mPi\vy_{\ii-1} + \ve_\ii.
\end{align}
In particular, the authors focus on the case where $\mathbf{v}_1 = \mfzero$.

Although model (\ref{eqn:model_h}) is equivalent to model (\ref{eqn:model_1})--(\ref{eqn:model_2}) under the parameter constraints $\mathbf{v} = -\mPi\ploc + (\id_{\p} + \mPi)\ptrd$ and $\mathbf{v}1 = -\mPi\ptrd$, it may generate quadratic time trends without these constraints. That being said, model (\ref{eqn:model_1})--(\ref{eqn:model_2}) has the advantage of emphasizing that the time trend considered in $\vy_\ii$ is at most linear (see \cite{lutkepohl2000testing} for further discussion). These different time trend specifications lead to distinct asymptotic results. Under model (\ref{eqn:model_h}), the cointegration test of \cite{hallin2016semiparametric} will have asymptotic power that depends on $\mathbf{v}$. Specifically, the test is more powerful when $\mathbf{v}$ is larger but has low power when the time trend is close to zero. In contrast, our test's asymptotic power does not depend on $\ploc$ or $\ptrd$, as they are eliminated using the invariance principle. 
\end{remark}

\section{Semiparametric Power Envelopes} \label{sec:semipowerenvelope}

\subsection{Preliminaries}
Our asymptotic analysis relies on the limit experiment approach. The limit experiment of a sequence of experiments (in this case, cointegration experiments) is defined by the convergence of likelihood ratios under specific local perturbations. To ensure that the likelihood-ratio convergence is neither explosive nor degenerate, we need to localize $\ploc$, $\ptrd$, $\mPi$ and $\f$ with appropriate rates. Such local perturbations are known as the \textit{contiguous} alternative (see \citet[Chapter 6]{vdVaart00}). In what follows, we introduce these local reparameterizations separately.

We follow the unit root and cointegration literature in adopting the \textit{local-to-unity} asymptotics for the key parameter of interest, $\mPi$. The associated local reparameterization is given by 
\begin{align} \label{eqn:ptb_mPi}
\mPi = \mPi^{(\n)}_{\mfC} = \frac{\mfC}{\n},
\end{align}
where $\mfC\in\R^{\p\times\p}$ is referred to the local parameter. The ``super-consistency'' contiguity rate $\n^{-1}$ here is common in models for nonstationary time series (see \citet{Phillips1987} and \cite{ChanWei88}), including unit root testing, cointegration, and predictive regression with persistent predictors.\footnote{For related predictive regression literature, see, e.g., \cite{elliott1994inference}, \citet{jansson2006optimal}, and \citet{werker2022semiparametric}.} This nonstandard rate $\n^{-1}$ leads to LABF-type experiments, as we show in Proposition~\ref{prop:limitexperiment_LABF} below.

We localize the time trend parameter as follows:
\begin{align} \label{eqn:ptb_trend}
\ptrd = \ptrd^{(\n)}_{\lptrd} = \ptrd_0 + \frac{\lptrd}{\sqrt{\n}},
\end{align}
where $\ptrd_0$ represents the true value. This local reparameterization is the natural multivariate counterpart of the unit root testing problem, as discussed in \citet[Section 7]{Jansson2008}. Following that paper, without loss of generality, we assume that $\ptrd_0$ is zero.

We do not localize the constant term parameter $\ploc$ in our approach since we show in the proof of Proposition~\ref{prop:limitexperiment_LABF} that it vanishes asymptotically. This result is in line with the findings of \citet[Section 7]{Jansson2008}, as the univariate counterpart, and show that information about $\ploc$ only comes from the first few observations rather than the data flow.

Finally, we adopt the approach by \citet{zhou2019semiparametrically} and introduce explicit nonparametric local perturbations to the innovation density $f$ as follows:
\begin{align}\label{eqn:ptb_density}
\f_{\lpptb}^{(\n)}(\ve) 
:=
\f(\ve)\left(1 + \frac{\lpptb^\trans}{\sqrt{\n}}\vfptb(\ve)\right).
\end{align} 
Here, $\vfptb := (\fptb_1,\fptb_2,\dots)^\trans$ is a vector of functions that govern the perturbations in different directions, and $\lpptb := (\eta_1,\eta_2,\dots)^\trans$ is the local parameter that determines the severity of these perturbations. We choose $\vfptb$ to be a countable orthonormal basis of the separable Hilbert space
\begin{align*}
\L_2^{0,\f}(\R^\p,\mathcal{B}) := \left\{\fptb\in\L_2^\f(\R^\p,\mathcal{B})\,\left|\,\E_\f[\fptb(\ve)]=0,\,\E_\f[\ve\fptb(\ve)]=0\right.\right\},
\end{align*}
$\ve\in\R^{\p}$, where $ \L_2^\f(\R^\p,\mathcal{B})$ denotes the space of Borel-measurable functions $\fptb:\R^\p\to\R$ that are square-integrable. Accordingly, we have $\var_\f[b_k(\ve)] = 1$. The separability of the Hilbert space ensures the existence of such a countable orthonormal basis. We further assume that $b_k \in C_{2,b}(\R)$ for all $k$, meaning that $b_k$'s are bounded and twice continuously differentiable with bounded derivatives.

We restrict the local perturbation parameter $\lpptb$ to have only finitely many non-zero elements, i.e., $\lpptb \in c_{00}$ where $c_{00} := \{(z_k)_{k\in\N}\in\R^{\N}\,|\,\sum_{k=1}^\infty \indicator\{z_k \neq 0\} < \infty\}$, which is a dense subspace of the parameter space $\ell_2 = \{(z_k)_{k\in\N}\,|\,\sum_{k=1}^\infty z_k^ 2<\infty\}$. This restriction does not sacrifice generality but rather helps avoid dealing with the convergence of infinite-dimensional Brownian motions and possibly associated measurability complexities. Notably, when $\lpptb = \mfzero$, we obtain $\f_{\lpptb}^{(\n)}=\f$. We demonstrate in the following proposition that for any $\lpptb \neq \mfzero$, $\f_{\lpptb}^{(\n)}$ satisfies Assumption~\ref{assm:density_fp}. The proof is detailed in Appendix~\ref{I-appendix:mainproofs}.

\smallskip
\begin{proposition} \label{prop:density_perturbed}
For any fixed $\f\in\F_\p$ and $\lpptb\in c_{00}$, there exists $\n^\prime \in \N$ such that $\f_{\lpptb}^{(\n)}\in\F_\p$ for all $\n\geq\n^\prime$. 
\end{proposition}
\smallskip

However, we do not aim to demonstrate that the particular nonparametric form $\f_{\lpptb}^{(\n)}$ with $\lpptb\in c_{00}$ is capable of generating all possible local perturbations on $\f$ in $\F_\p$. Nonetheless, this concrete perturbation specification does not affect our analysis of semiparametric optimality, particularly the derivation of the semiparametric power envelopes. Essentially, $\f_{\lpptb}^{(\n)}$ can be regarded as a complex sub-model under which a local power upper bound can be derived. Once a feasible semiparametric test is constructed that can attain this upper bound (as we will demonstrate in Section~\ref{sec:semi_inference}), it becomes the semiparametric power envelope.

\subsection{The limiting experiment}
We define $\law_{\mfC,\lptrd,\lpptb;\ploc,\f}$ as the law of $\vy_1,\dots,\vy_\n$ generated by the error-correction model (\ref{eqn:model_1})--(\ref{eqn:model_2}) with local reparameterizations in (\ref{eqn:ptb_mPi})--(\ref{eqn:ptb_density}). Additionally, we introduce the probability measure of the associated limit experiment, denoted as $\prob_{\mfC,\lptrd,\lpptb}$, which will be formally introduced later. In the following proposition, we demonstrate that the log-likelihood ratio process $\LR^{(\n)}_\f(\mfC,\lptrd,\lpptb):= \log\big(\dd\law_{\mfC,\lptrd,\lpptb;\ploc,\f}/\dd\law_{\mfzero,\mfzero,\mfzero;\ploc,\f}\big)$ follows the Locally Asymptotically Brownian Functional form introduced by \cite{Jeganathan1995}.

\smallskip 
\begin{proposition} \label{prop:limitexperiment_LABF}
Consider $\f\in\F_\p$. Let $\mfC\in\R^{p\times p}$, $\ploc\in\R^{\p}$, $\lptrd\in\R^{\p}$, and $\lpptb\in c_{00}$.
\begin{itemize}
\item[(a)] 
Under $\law_{\mfzero,\mfzero,\mfzero;\ploc,\f}$, as $\n\to\infty$, the log-likelihood ratio is decomposed as 
\begin{align} \label{eqn:LR_sequence}
\LR^{(\n)}_\f(\mfC,\lptrd,\lpptb) = \CS^{(\n)}_\f(\mfC,\lptrd,\lpptb)-\frac{1}{2}\QV^{(\n)}_\f(\mfC,\lptrd,\lpptb)+o_{P}(1),  
\end{align}
where
\begin{align*}
\CS^{(\n)}_\f(\mfC,\lptrd,\lpptb) 
:=&~ 
\frac{1}{\n}\sum_{\ii=2}^{\n}(\mfC\vyy)^{\trans}\score_{f}(\Delta\vy_\ii) + \frac{1}{\sqrt{\n}}\sum_{\ii=2}^{\n}\left(\dct\lptrd\right)^\trans\score_\f(\Delta \vy_\ii) + \frac{1}{\sqrt{\n}}\sum_{\ii=2}^{\n}\lpptb^{\trans}\vfptb(\Delta \vy_\ii), 
\\
\QV^{(\n)}_\f(\mfC,\lptrd,\lpptb) 
:=&~ 
\frac{1}{\n^2}\sum_{\ii=2}^{\n}(\mfC\vyy)^{\trans}\J_\f\mfC\vyy
+ \frac{1}{\n}\sum_{\ii=2}^{\n}(\dct\lptrd)^{\trans}\J_\f\dct\lptrd \\
&~ + \frac{2}{\n^{3/2}}\sum_{\ii=2}^{\n}(\mfC\vyy)^{\trans}\J_\f\dct\lptrd + \frac{2}{T^{3/2}}\sum_{\ii=2}^{\n}\lpptb^{\trans} \J_{\vfptb\f}\mfC\vyy \\
&~ + \frac{2}{T}\sum_{\ii=2}^{\n}\lpptb^{\trans} \J_{\vfptb\f}\dct\lptrd + \lpptb^{\trans}\lpptb,
\end{align*}
with $\dct := \id_{\p}-\frac{\ii-1}{\n}\mfC$, $\vyy := \vy_{t-1} - \vy_{1}$, and $\J_{\vfptb\f} := \E_{\f}[\vfptb(\ve_1)\score_\f(\ve_1)^\trans]$.
\item[(b)] 
Let $\We,\Wlf,\Wb$ be Brownian motions defined on the probability space $(\Omega,\mathcal{F},\prob_{\mfzero,\mfzero,\mfzero})$ with covariance 
\begin{align} \label{eqn:covariancematrix}
\var\begin{pmatrix}\We(1)\\\Wlf(1)\\\Wb(1)\end{pmatrix}=
\begin{pmatrix}
\Sig & \id_{\p} & \mfzero_{\p,\infty} \\
\id_{\p} & \J_{\f} & \J_{\f\vfptb} \\
\mfzero_{\infty,\p} & \J_{\vfptb\f} & \id_\infty
\end{pmatrix},
\end{align}
where $\id$ and $\mfzero$ are the identity and zero matrices of dimensions shown in their superscripts, respectively. Then, under $\law_{\mfzero,\mfzero,\mfzero;\ploc,\f}$ and as $\n\to\infty$, we have 
\begin{align} \label{eqn:limitlikelihood}
\LR^{(\n)}_\f(\mfC,\lptrd,\lpptb)\wto\LR_\f(\mfC,\lptrd,\lpptb) = \CS_\f(\mfC,\lptrd,\lpptb)-\frac{1}{2}\QV_\f(\mfC,\lptrd,\lpptb),
\end{align}
where
\begin{align*}
\CS_\f(\mfC,\lptrd,\lpptb) :=& \int_0^1(\mfC\We(\s))^{\trans}\dd\Wlf(\s) + \int_0^1(\dcu\lptrd)^{\trans}\dd\Wlf(\s)+\lpptb^{\trans}\Wb(1), \\
\QV_\f(\mfC,\lptrd,\lpptb) :=& \int_0^1(\mfC\We(\s))^{\trans}\J_\f C\We(\s)\dd\s + \int_0^1(\dcu\lptrd)^{\trans}\J_\f\dcu\lptrd\dd\s \\
&~ + 2\int_0^1(\mfC\We(\s))^{\trans}\J_\f\dcu\lptrd\dd\s + 2\lpptb^{\trans}\J_{\vfptb\f}\mfC\widebar\W_{\ve} + 2\lpptb^{\trans}\J_{\vfptb\f}\dcub\lptrd + \lpptb^\trans\lpptb,
\end{align*}
with $\dcu := \id_{\p} - \s\mfC$, $\dcub := \int_0^1\dcu\dd\s = \id_{\p} - \mfC/2$, and $\widebar\W_{\ve} := \int_0^1\We(\s)\dd\s$.
\item[(c)] 
Under $\prob_{\mfzero,\mfzero,\mfzero}$, $\forall \, \mfC\in\R^{p \times p}$, $\lptrd\in\R^{\p}$, and $\lpptb\in c_{00}$, $\Elim\left[\exp(\LR_\f(\mfC,\lptrd,\lpptb))\right] = 1$.
\end{itemize}
\end{proposition}
\smallskip

The proof of part (a) could have essentially followed from a Taylor expansion, assuming twice continuous differentiability on $\f$. However, instead, we employed the framework of \citet[Proposition 1]{hallin2015quadratic}, which is built upon the DQM condition and is implied by the absolute continuity assumption. This allows for a broader family of innovation density $f$, including distributions such the double exponential distribution. The proof for Part (b) is based on the functional central limit theorem, the continuous mapping theorem, and an application of \citet[Theorem 2.1]{hansen1992convergence}. Part (c) follows from standard stochastic calculus of the Dol\'eans-Dade exponential, once the Novikov's condition is verified. The detailed proofs are organized in Appendix~\ref{I-appendix:mainproofs} of the supplementary material.

Part (a) and Part (b) demonstrate that the limit experiment, specifically with respect to $\mPi$, is LABF as defined by \citet{Jeganathan1995}. In other words, the central sequence weakly converges to a stochastic integral where the integrand and integrator processes exhibit correlation. This is evident from the covariance matrix equation (\ref{eqn:covariancematrix}), which shows that $\cov(\We(1),\Wlf(1)) = \id_\p$.

Part (c) allows us to introduce a new collection of probability measures, denoted by $\prob_{\mfC,\lptrd,\lpptb}$, through its Radon-Nikodym derivative w.r.t.\ $\prob_{\mfzero,\mfzero,\mfzero}$, given by
\begin{align*}
\frac{\dd\prob_{\mfC,\lptrd,\lpptb}}{\dd\prob_{\mfzero,\mfzero,\mfzero}} = \exp\LR_\f(\mfC,\lptrd,\lpptb).
\end{align*}
The measurable space $(\Omega,\mathcal{F})$ is that of the Brownian motions $(\We,\Wlf,\Wb)$, which are defined as $\Omega := \C^\p[0,1]\times\C^\p[0,1]\times\C^{\infty}[0,1]$ and $\filtration := (\otimes_\p\borel_\C) \otimes (\otimes_\p\borel_\C) \otimes (\otimes_{k=1}^{\infty}\borel_\C)$, where $\borel_\C$ denotes the Borel $\sigma$-field on $\C[0,1]$.

Having introduced the necessary ingredients, we can now provide a formal definition of the limit experiment as 
\begin{align*} 
\experiment(\f) := \left(\Omega, \filtration, \left\{\prob_{\mfC,\lptrd,\lpptb}: \mfC\in\R^{\p\times\p}, \lptrd\in\R^\p, \lpptb\in c_{00} \right\} \right).
\end{align*}
Then, in Le Cam's sense (see, e.g., \citet[Chapter 9]{vdVaart00}), the sequence of cointegration experiments, denoted by $\experiment^{(\n)}(\f)$, converges to the limit experiment $\experiment(\f)$ as the sample size $\n$ tends to infinity. In the following proposition, we regard $\exp\LR_\f(\mfC,\lptrd,\lpptb)$ as the Radon-Nikodym derivative and apply Girsanov's Theorem to obtain a structural representation of the limit experiment $\experiment(\f)$.

\smallskip
\begin{proposition} \label{prop:structural_representation}
Let $\mfC\in\R^{\p\times\p}$, $\lptrd\in\R^{\p}$, $\lpptb\in c_{00}$, and fix $\f\in\F_{\p}$. The limit experiment $\experiment(\f)$ associated with the log-likelihood ratio $\LR_\f(\mfC,\lptrd,\lpptb)$ can be described as follows. We observe processes $\We$, $\Wlf$ and $\Wb$, which are generated according to the following stochastic differential equations (SDEs):
\begin{align}
\dd\We(\s)  =&~ \mfC\We(\s)\dd\s + \dcu\lptrd\dd\s + \dd\Ze(\s), \label{eqn:We} \\
\dd\Wlf(\s) =&~ \J_\f \mfC\We(\s)\dd\s + \J_\f\dcu\lptrd\dd\s + \J_{\f\vfptb}\lpptb\dd\s + \dd\Zlf(\s), \label{eqn:Wlf} \\
\dd\Wb(\s)  =&~ \J_{\vfptb\f} \mfC\We(\s)\dd\s + \J_{\vfptb\f}\dcu\lptrd\dd\s  + \lpptb\dd\s + \dd\Zb(\s), \label{eqn:Wb}
\end{align}
where $\Ze$, $\Zlf$ and $\Zb$ are Brownian motions under $\prob_{\mfC,\lptrd,\lpptb}$ with drift zero and covariance given by (\ref{eqn:covariancematrix}). 
\end{proposition}
\smallskip

Next, we will use this structural limit experiment to eliminate the nuisance parameters, $\lpptb$ and $\lptrd$, sequentially. By doing so, we will derive the semiparametric power envelopes for both cases, without and with a time trend.

\subsection{\texorpdfstring{Eliminating $\lpptb$}{} using Brownian bridge} \label{subsec:eliminate_lpptb}
Despite exhibiting LABF behavior with respect to the direction of $\mfC$, the limit experiment remains LAN with respect to $\lpptb$. This is a critical feature, as it results in that $\lpptb$ only appears as constant drifts in (\ref{eqn:We})--(\ref{eqn:Wb}). To eliminate these drifts, we can simply ``take the bridges'' of the affected processes.

We formally define the transformation $\mkg_{\lpptb}$ as follows:
\begin{align} \label{eqn:transformation_g}
\mkg_{\lpptb} \, : \, [\mkg_{\lpptb}(\W)](\s) = \W(\s) - \lpptb\s, ~~~ \s\in[0,1]
\end{align}
for a process $\W\in D^{\N}[0,1]$. We denote by $\mkG_{\lpptb}$ the group of $\mkg_{\lpptb}$ for $\lpptb\in c_{00}$. Intuitively, the transformation $\mkg_{\lpptb}$ adds a constant drift $\s \to -\lpptb\s$ to $\W$. To eliminate such a constant drift, we employ the bridge-taking operator defined as
\begin{align} \label{eqn:bridge}
\B^{\W}(\s) := \W(\s) - \s\W(1).
\end{align}
For a fixed $\lpptb\in c_{00}$, we have $\B^{[\mkg_{\lpptb}(\W)]}(\s) = [\mkg_{\lpptb}(\W)](\s) - \s[\mkg_{\lpptb}(\W)](1) = (\W(\s) - \lpptb\s) - \s(\W(1) - \lpptb) = \W(\s) - \s\W(1) = \B^{\W}(\s)$. This shows that the deduced bridge process is an invariant statistic with respect to $\mkG_{\lpptb}$.

Note that in the structural limit experiment described in (\ref{eqn:We})--(\ref{eqn:Wb}), the parameter $\lpptb$ appears in $\Wlf$ and $\Wb$, but not in $\We$.  Therefore, to eliminate the constant drifts caused by $\lpptb$, we take the bridges of $\Wlf$ and $\Wb$ while keeping $\We$ unchanged, resulting in an invariant $\sigma$-field given by
\begin{align} \label{eqn:mi}
\mi := \sigma\big(\We,\Blf,\Bb\big),
\end{align}
where $\Blf := \B^{\Wlf}$ and $\Bb := \B^{\Wb}$. Furthermore, we show below that $\mi$ is \textit{maximally invariant}.

\smallskip
\begin{theorem} \label{thm:maximalinvariant}
In the limit experiment $\experiment(\f)$ modeled by (\ref{eqn:We})--(\ref{eqn:Wb}), the $\sigma$-field $\mi$ defined in (\ref{eqn:mi}) is maximally invariant with respect to the group of transformations $\mkG_{\lpptb}$, where $\lpptb\in c_{00}$.
\end{theorem}
\smallskip

The proof of Theorem~\ref{thm:maximalinvariant} is based on the definition of \textit{maximal invariant} in Section 6.2 of \citet{LehmannRomano2005}.\footnote{Note that $\sigma(\We,\Wlf,\Wb) = \sigma(\We,\Wb)$ due to the decomposition $\Wlf = \Sig^{-1}\We + \J_{\f\vfptb}\Wb$ according to the covariance (\ref{eqn:covariancematrix}). This fact simplifies the proof.} The detailed proof is provided in Appendix~\ref{I-appendix:mainproofs} of the supplementary material. The maximal invariant result of $\mi$ plays the vital role in our semiparametric optimality study, as every invariant statistic (w.r.t.\ $\lpptb$) is $\mi$-measurable (see \citet[Theorem 6.2.1]{LehmannRomano2005}), and is every invariant test (w.r.t.\ $\lpptb$). Therefore, by the Neyman-Pearson Lemma, the likelihood ratio test based on $\mi$ is optimal among all invariant test (w.r.t.\ $\lpptb$). The log-likelihood ratio of $\mi$ is given by
\begin{align} \label{eqn:likelihood_MI}
\LR^{\mi}_\f(\mfC,\lptrd) = \CS^{\mi}_\f(\mfC,\lptrd)-\frac{1}{2}\QV^{\mi}_\f(\mfC,\lptrd),
\end{align} 
where 
\begin{align*}
\CS^{\mi}_\f(\mfC,\lptrd) =& \int_0^1\left(\mfC\We(\s)+\dcu\lptrd\right)^\trans\dd\big(\Blf(\s)+\Sig^{-1}\We(1)\s\big) \\
\QV^{\mi}_\f(\mfC,\lptrd) =& \int_0^1\left(\mfC\We(\s)+\dcu\lptrd\right)^\trans\J_\f(\mfC\We(\s)+\dcu\lptrd)\dd\s \\
&~ + \left(\mfC\widebar\W_{\ve}+\dcub\lptrd\right)^\trans\left(\Sig^{-1}-\J_\f\right)\left(\mfC\widebar\W_{\ve}+\dcub\lptrd\right).
\end{align*}
The detailed derivation is provided in the supplementary Appendix~\ref{I-appendix:calculation}.

Up to this point, we have applied the invariance principle to eliminate the nuisance parameter $\lpptb$, leaving us with another nuisance parameter $\lptrd$ to address. In the following subsections, we will sequentially consider two cases: one in which there is no time trend ($\lptrd = 0$), and another in which there is a time trend ($\lptrd$ is unknown).

\subsection{Semiparametric power envelope, no-time-trend case}
We begin by considering the case of an intercept only, which corresponds to the error-correction model in (\ref{eqn:model_1})--(\ref{eqn:model_2}) with $\ptrd = \mfzero$ (or equivalently, $\lptrd = \mfzero$). This specification is important in its own right and is perhaps the most commonly used model in applied works. The advantage of the no-time-trend specification is that the associated tests enjoy better power performance, albeit with a potential cost of model misspecification, when researchers do not find obvious evidence of a time trend in their datasets. For further discussions on this issue, we refer to \cite{lutkepohl2000testing} and \cite{saikkonen2000trend}.

We use $\LR^{\ploc*}_\f$ to denote the log-likelihood ratio associated with the maximal invariant under $\lptrd = \mfzero$, for which we have 
\begin{align} \label{eqn:likelihood_MI_mu}
\LR^{\ploc*}_\f(\bar\mfC) := \LR^{\mi}_\f(\bar\mfC,\mfzero) = \CS^{\ploc*}_\f(\bar\mfC)-\frac{1}{2}\QV^{\ploc*}_\f(\bar\mfC),
\end{align} 
where
\begin{align*}
\CS^{\ploc*}_\f(\bar\mfC) :=& \int_0^1\left(\bar\mfC\We(\s)\right)^{\trans}\dd\big(\Blf(\s) + \Sig^{-1}\We(1)\s\big)  \\
\QV^{\ploc*}_\f(\bar\mfC) :=& \int_0^1\left(\bar\mfC\We(\s)\right)^{\trans}\J_\f \bar\mfC\We(\s)\dd\s + \left(\bar\mfC\widebar\W_{\ve}\right)^{\trans}\left(\Sig^{-1}-\J_\f\right)\bar\mfC\widebar\W_{\ve}.
\end{align*}
We define the associated likelihood ratio test by $\test^{\ploc*}_{\f,\alpha}(\bar\mfC) := \indicator\{\LR^{\ploc*}_\f(\bar\mfC) > \cv^{\ploc}_\alpha(\bar\mfC)\}$, where $\cv^{\ploc}_\alpha(\bar\mfC)$ is the $1-\alpha$ quantile of $\LR^{\ploc*}_\f(\bar\mfC)$ under $\prob_{\mfzero,\mfzero,\lpptb}$. An upper power bound for $\lpptb$-invariant tests can then be given as follows:
\begin{align*}
\mfC
\mapsto
\power^{\ploc*}_{\f,\alpha}(\bar\mfC;\mfC) = \Elim\left[\test^{\ploc*}_{\f,\alpha}(\bar\mfC) \frac{\dd\prob_{\mfC,\mfzero,\lpptb}}{\dd\prob_{\mfzero,\mfzero,\mfzero}}\right].
\end{align*}

Call a test $\testT$ in $\experiment^{(\n)}(\f)$ \textit{asymptotically $\lpptb$-invariant} if it weakly converges, under $\law_{\mfC,\mfzero,\lpptb;\ploc,\f}$, to a test $\test$ in $\experiment(\f)$ that is invariant w.r.t.\ $\lpptb$. The combination of the Neyman-Pearson Lemma and the Asymptotic Representation Theorem (\citet[Chapter 9]{vdVaart00}) yields the following theorem.

\smallskip
\begin{theorem} \label{thm:powerenvelope_mu}
Assuming that $\lptrd = \mfzero$ is known, let $\f\in\F_{\p}$, $\ploc\in\R^{\p}$, and $\alpha\in(0,1)$. If a test $\testT(\vy_1,\dots,\vy_\n)$, where $\n\in\N$, is an asymptotically $\lpptb$-invariant test of size $\alpha$, i.e., $\limsup_{\n\to\infty}\E_{\mfzero,\mfzero,\lpptb;\ploc,\f}[\testT]\leq\alpha$, then we have
\begin{align*}
\limsup_{\n\to\infty} \E_{\mfC,\mfzero,\lpptb;\ploc,\f}[\testT] 
\leq 
\power^{\ploc*}_{\f,\alpha}(\mfC;\mfC), ~~~\forall \, \mfC\in\R^{\p\times\p}, \, \lpptb\in c_{00}.
\end{align*}
\end{theorem}
\smallskip

The proof of the theorem involves two main steps. First, we use the Neyman-Pearson lemma to establish the upper bound for the power of $\lpptb$-invariant tests in $\experiment(\f)$ at point $\mfC$, which is $\power^{\ploc*}_{\f,\alpha}(\mfC;\mfC)$. This determines the maximum achievable power in the limit experiment. Second, the Asymptotic Representation Theorem states that any test in the sequence of experiments $\experiment^{(\n)}(\f)$ has a representation in the limit experiment $\experiment(\f)$. Therefore, the best achievable power in $\experiment(\f)$ is also the best possible power that can be achieved asymptotically in $\experiment^{(\n)}(\f)$. The proof is completed by applying this result to the class of asymptotically $\lpptb$-invariant tests.

\subsection{\texorpdfstring{Eliminating $\lptrd$ using profile likelihood \& semiparametric power envelope for time-trend case}{}} \label{subsec:eliminate_lptrd}
In this subsection, we consider case where the linear trend parameter $\lptrd$ is unknown and treated as a nuisance parameter. To eliminate $\lptrd$, observing that the log-likelihood ratio $\LR^{\mi}_\f(\mfC,\lptrd)$ is quadratic in $\lptrd$, we use the \textit{profile likelihood} method. Specifically, $\lptrd$ is ``profiled out'' by 
\begin{align} \label{eqn:profilelikelihood}
\max_{\lptrd}\LR^{\mi}_\f(\mfC,\lptrd) - \max_{\lptrd}\LR^{\mi}_\f(\mfzero,\lptrd),
\end{align}
which results in an invariant statistic w.r.t $\lptrd$.\footnote{More formally, the obtained statistic is invariant w.r.t.\ the group of transformations of the form $\vy_t = \vy_t + \mfc\s$, $\forall\mfc\in\R^{\p}$.} The optimality of this approach for quadratic-in-nuisance form is established in, e.g., \citet[Problem 6.9]{LehmannRomano2005}, where it is shown that the resulting profile likelihood-based test is the best among invariant tests. For further reference on this method in the unit root testing literature, see \citet{elliott1996efficient} and \citet{Jansson2008}.

We split the optimization of (\ref{eqn:profilelikelihood}) into two steps: (i) we derive the maximum likelihood estimate of $\lptrd$ by solving the maximization under one alternative value of $\mfC$, and (ii) we plug this estimate into likelihood statistic $\LR^{\mi}_\f$ under another alternative value. This splitting allows us to obtain the semiparametric analogue of the statistic $\mathit{\Lambda}_{p,C}^{GLS}(\bar{C};\bar{C}^*)$ introduced in \citet[Section 2.2]{boswijk2015improved} (BJN). This statistic encompasses existing Gaussian cointegration test statistics obtained by choosing different values of $\bar{C}$ and $\bar{C}^*$. By interpreting the choices of $\bar{C}$ and $\bar{C}^*$ in BJN's $\mathit{\Lambda}_{p,C}^{GLS}(\bar{C};\bar{C}^*)$, we contribute to the literature by providing insights into how the current Gaussian cointegration tests for the time trend case are derived in the limiting perspective, and by proposing semiparametric efficient versions of these tests using our semiparametric analogue afterwards.

In detail, we select the reference alternative $\bar\mfC^*$ and derive the maximum likelihood estimate (MLE) of $\lptrd$ as follows:
\begin{align} \label{eqn:lptrd_estimate}
\hat{\lptrd}_{\bar\mfC^*} 
&= {\rm arg\,max}_{\lptrd\in\R^{\p}}\LR^{\mi}_\f(\bar\mfC^*,\lptrd) \nonumber \\
&= \mathbf{B}(\bar\mfC^*)^{-1}\mathbf{A}(\bar\mfC^*),
\end{align}
where 
\begin{align*}
\mathbf{A}(\mfC) :=&~ \int_0^1\dcu^{\trans}\dd\big(\Blf(\s)+\Sig^{-1}\We(1)\s - \J_\f\mfC\widetilde{\W}_{\ve}(\s)\s - \Sig^{-1}\mfC\widebar{\W}_{\ve}\s\big), \\
\mathbf{B}(\mfC) :=&~ \int_0^1\dcu^\trans\J_\f\dcu\dd\s + \widebar{\mfd}_{\mfC}^\trans\left(\Sig^{-1}-\J_\f\right)\widebar{\mfd}_{\mfC}
\end{align*}
are the linear and quadratic term in $\lptrd$, respectively. Once we have $\hat\lptrd_{\bar\mfC^*}$, we use another reference alternative $\bar\mfC$ to construct the profile likelihood ratio as
\begin{align} \label{eqn:likelihood_MI_tau}
\LR^{\ptrd*}_\f(\bar\mfC;\bar\mfC^*)
&= \LR^{\mi}_\f(\bar\mfC,\hat{\lptrd}_{\bar\mfC^*}) - \LR^{\mi}_\f(\mfzero,\hat\lptrd_{\mfzero}) \nonumber \\
&= \CS^{\ptrd*}_\f(\bar\mfC;\bar\mfC^*) - \frac{1}{2}\QV^{\ptrd*}_\f(\bar\mfC;\bar\mfC^*) - \frac{1}{2}\W^{\hat\lptrd_{\bar\mfC^*}}_{\ve}(1)^\trans\Sig^{-1}\W^{\hat\lptrd_{\bar\mfC^*}}_{\ve}(1)
\end{align}
with
\begin{align*}
\CS^{\ptrd*}_\f(\bar\mfC;\bar\mfC^*) :=&~ \int_0^1\big(\bar\mfC\W_{\ve}^{\hat\lptrd_{\bar\mfC^*}}(\s)\big)^\trans\dd\big(\Blf(\s)+\Sig^{-1}\W^{\hat\lptrd_{\bar\mfC^*}}_{\ve}(1)\s\big), \\
\QV^{\ptrd*}_\f(\bar\mfC;\bar\mfC^*) :=&~ 
\int_0^1\big(\bar\mfC\W_{\ve}^{\hat\lptrd_{\bar\mfC^*}}(\s)\big)^\trans\J_\f\bar\mfC\W_{\ve}^{\hat\lptrd_{\bar\mfC^*}}(\s)\dd\s  \\
&~ + \big(\bar\mfC\widebar\W_{\ve}^{\hat\lptrd_{\bar\mfC^*}}\big)^\trans\big(\Sig^{-1}-\J_\f\big)\bar\mfC\widebar\W_{\ve}^{\hat\lptrd_{\bar\mfC^*}},
\end{align*}
where $\W_{\ve}^{\hat\lptrd_{\bar\mfC^*}}(\s) := \We(\s) - \s\hat\lptrd_{\bar\mfC^*}$ is the de-drifted version (of $\We$) with drift estimate $\hat\lptrd_{\bar\mfC^*}$, and $\widebar\W_{\ve}^{\hat\lptrd_{\bar\mfC^*}} := \widebar\We - \frac{1}{2}\hat\lptrd_{\bar\mfC^*}$ is its average.\footnote{Here the profile log-likelihood $\LR^{\mi}_\f(\bar\mfC,\hat{\lptrd}_{\bar\mfC^*}) - \LR^{\mi}_\f(\mfzero,\hat\lptrd_{\mfzero})$ is derived by a sum of two terms, $\LR^{\mi}_\f(\mfzero,\hat{\lptrd}_{\bar\mfC^*}) - \LR^{\mi}_\f(\mfzero,\hat\lptrd_{\mfzero}) = -0.5\,\W^{\hat\lptrd_{\bar\mfC^*}}_{\ve}(1)^\trans\Sig^{-1}\W^{\hat\lptrd_{\bar\mfC^*}}_{\ve}(1)$ and $\LR^{\mi}_\f(\bar\mfC,\hat{\lptrd}_{\bar\mfC^*}) - \LR^{\mi}_\f(\mfzero,\hat\lptrd_{\bar\mfC^*}) = \CS^{\ptrd*}_\f(\bar\mfC;\bar\mfC^*) - \frac{1}{2}\QV^{\ptrd*}_\f(\bar\mfC;\bar\mfC^*)$, where $\hat\lptrd_{\mfzero}$ denotes the estimate for $\lptrd$ under the chosen alternative $\bar\mfC^* = \mfzero$. By (\ref{eqn:lptrd_estimate}), we have $\hat\lptrd_{\mfzero} = \We(1)$. The second term is intuitive and easy-to-interpret --- it is nothing but the likelihood ratio of the sigma-field $\sigma(\W_{\ve}^{\hat\lptrd_{\bar\mfC^*}}, \Blf)$.}

Our statistic $\LR^{\ptrd*}\f(\bar\mfC;\bar\mfC^*)$ serves as the semiparametric counterpart and reduces to BJN's $\mathit{\Lambda}{p,C}^{GLS}(\bar{C};\bar{C}^*)$ when $\f$ is Gaussian. Specifically, under Gaussianity, the score function is given by $\score_\f(\ve_\ii) = \Sigma^{-1}\ve_\ii$, and the limit experiment in Proposition~\ref{prop:structural_representation} simplifies to (\ref{eqn:We}). Using $\bar\mfC^*$ to estimate $\lptrd$ and $\bar\mfC$ to construct the statistic, we arrive at BJN's $\mathit{\Lambda}_{p,C}^{GLS}(\bar{C};\bar{C}^*)$. It is worth noting again that $\mathit{\Lambda}{p,C}^{GLS}(\bar{C};\bar{C}^*)$ embodies asymptotic representations of existing Gaussian cointegration tests. In a similar vein, in Section~\ref{sec:semi_inference}, we will base on our $\LR^{\ptrd*}_\f(\bar\mfC;\bar\mfC^*)$ to derive their semiparametric versions.

We conclude this section by presenting the semiparametric power envelope for the case of a linear time trend. We let $\bar\mfC^* = \bar\mfC$ and define the likelihood ratio test as $\test^{\ptrd*}_{\f,\alpha}(\bar\mfC) := \indicator\{\LR^{\ptrd*}_\f(\bar\mfC;\bar\mfC) > \cv^{\ptrd}_\alpha(\bar\mfC)\}$, where $\cv^{\ptrd}_\alpha(\bar\mfC)$ is the $1-\alpha$ quantile of $\LR^{\ptrd*}_\f(\bar\mfC;\bar\mfC)$. The power function is given by
\begin{align*}
\mfC
\mapsto
\power^{\ptrd*}_{\f,\alpha}(\bar\mfC;\mfC) = \Elim\left[\test^{\ptrd*}_{\f,\alpha}(\bar\mfC) \frac{\dd\prob_{\mfC,\lptrd,\lpptb}}{\dd\prob_{\mfzero,\mfzero,\mfzero}}\right].
\end{align*}
We refer to a test $\test_{\n}$ in $\experiment^{(\n)}(\f)$ as \textit{asymptotically $(\lptrd,\lpptb)$-invariant} if it weakly converges to a test in $\experiment(\f)$ that is invariant w.r.t.\ both $\lpptb$ and $\lptrd$. Using reasoning analogous to Theorem~\ref{thm:powerenvelope_mu}, we provide the semiparametric power envelope of asymptotically $(\lptrd,\lpptb)$-invariant tests for the time trend case in the following theorem.

\smallskip
\begin{theorem} \label{thm:powerenvelope_tau}
Assuming that $\lptrd$ is unknown, let $\f\in\F_{\p}$, $\ploc\in\R^{\p}$, and $\alpha\in(0,1)$. If a test $\testT(\vy_1,\dots,\vy_\n)$, where $\n\in\N$, is an asymptotically $(\lptrd,\lpptb)$-invariant test of size $\alpha$, that is, $\limsup_{\n\to\infty}\E_{\mfzero,\lptrd,\lpptb;\ploc,\f}[\test^{\n}]\leq\alpha$, then we have
\begin{align*}
\limsup_{\n\to\infty}\E_{\mfC,\lptrd,\lpptb;\ploc,\f}[\testT] 
\leq 
\power^{\ptrd*}_{\f,\alpha}(\mfC;\mfC), ~~~\forall \, \mfC\in\R^{\p\times\p}, \, \lptrd\in\R^\p, \, \lpptb\in c_{00}. 
\end{align*}
\end{theorem}
\smallskip

\section{Semiparametric Inference} \label{sec:semi_inference}
This section proposes semiparametrically optimal cointegration tests based on the asymptotic results above. We first employ $\LR^{\ploc*}_\f(\mfC)$ (resp.\ $\LR^{\ptrd*}_\f(\bar\mfC;\bar\mfC^*)$) to derive (asymptotic) semiparametric counterparts of some existing Gaussian cointegration tests for the case without time trend (resp.\ the case with time trend) in Section~\ref{subsec:linking_notimetrend} (resp.\ Section~\ref{subsec:linking_timetrend}). Then, in Section~\ref{subsec:test}, we construct feasible semiparametric cointegration tests primarily by replacing $\Blf$ by its finite-sample counterpart, which is built upon a nonparametric estimate of the score function $\score_\f$.

\subsection{Semiparametric counterparts of existing tests (no-time-trend case)} \label{subsec:linking_notimetrend}
When there is no time trend ($\lptrd = \mfzero$), we follow the line of likelihood ratio test by \citet{johansen1991estimation} and \citet{saikkonen1997testing}, and profile out the chosen alternative $\bar\mfC$. Specifically, we maximize the likelihood ratio statistic $\LR^{\ploc*}_\f(\bar\mfC)$ in (\ref{eqn:likelihood_MI_mu}) w.r.t.\ $\bar\mfC\in\R^{\p\times\p}$, and obtain the \textit{semiparametric Johansen trace statistic}
\begin{align} \label{eqn:stat_johansen_semi}
\Lambda^{Johansen}_\f := \trace\,&\left(\int_0^1 \We(\s)\dd\big(\Blf(\s) + \Sig^{-1}\We(1)\s\big)^{\trans}\right)^\trans \nonumber \\
&~ \times \left(\int_0^1\We(\s)\J_\f\We(\s)^{\trans}\dd\s + \widebar\W_{\ve}\left(\Sig^{-1}-\J_\f\right)\widebar\W_{\ve}\right)^{-1} \nonumber \\
&~ \times \left(\int_0^1 \We(\s)\dd\big(\Blf(\s) + \Sig^{-1}\We(1)\s\big)^{\trans}\right),
\end{align}
which does not require a specification for $\bar\mfC$ anymore. See \citet{jansson2012nearly} for a comprehensive study on this operation for unit root testing. Borrowing their conclusion to our multivariate case, we argue that the resulting $\Lambda^{Johansen}_\f$-based test attains $\power^{\ploc*}_{\f,\alpha}(\mfC;\mfC)$ pointwise at a random alternative point.

Under the Gaussianity of the true density, i.e., $\f = \phi$, we have $\Blf = \Sig^{-1}\Be$ and $\J_\f = \Sig^{-1}$, and the semiparametric statistic $\Lambda^{Johansen}_\f$ simplifies to the original \textit{Gaussian Johansen trace statistic} as follows: 
\begin{align} \label{eqn:stat_johansen_phi}
\Lambda^{Johansen}_{\phi} :=&~ \trace\,\left(\int_0^1 \We(\s)\Sig^{-1}\dd\We(\s)^{\trans}\right)^\trans \nonumber \\
&~ \times \left(\int_0^1\We(\s)\Sig^{-1}\We(\s)^{\trans}\dd\s\right)^{-1} \left(\int_0^1 \We(\s)\Sig^{-1}\dd\We(\s)^{\trans}\right).
\end{align}

\subsection{Semiparametric counterparts of existing tests (time-trend case)} \label{subsec:linking_timetrend}
When dealing with a linear time trend (i.e., unknown $\lptrd$), we consider the GLS trace test proposed by \cite{saikkonen2000trend} (SL test) (see also \cite{lutkepohl2000testing}). In this case, we choose $\bar\mfC^* = \mfzero$ for the trend parameter estimate $\hat\lptrd_{\bar\mfC^*}$, which leads to $\hat\lptrd_{\bar\mfC^*} = \hat\lptrd_{\mfzero} = \We(1)$. Consequently, $\W_{\ve}^{\hat\lptrd_{\bar\mfC^*}}(\s) = \We(\s) - \s\We(1) = \Be(\s)$ and $\W_{\ve}^{\hat\lptrd_{\bar\mfC^*}}(1) = \mfzero$. The likelihood ratio statistic is then given by:
\begin{align*} 
\LR^{\ptrd*}_\f(\bar\mfC;\mfzero)
= \int_0^1\big(\bar\mfC\Be(\s)\big)^\trans\dd\Blf(\s) - \frac{1}{2}\int_0^1\big(\bar\mfC\Be(\s)\big)^\trans\J_\f\big(\bar\mfC\Be(\s)\big)\dd\s. 
\end{align*}
By maximizing $\LR^{\ptrd*}_\f(\bar\mfC;\mfzero)$ w.r.t.\ $\bar\mfC$, we obtain the \textit{semiparametric SL trace statistic} defined as
\begin{align} \label{eqn:stat_SL_semi}
\Lambda^{SL}_\f \,:= \, &\trace\,\left(\int_0^1\Be(\s)\dd\Blf(\s)^\trans\right)^\trans \nonumber \\
&~~~~~ \times \left(\int_0^1\Be(\s)\J_\f\Be(\s)^\trans\dd\s\right)^{-1}\left(\int_0^1\Be(\s)\dd\Blf(\s)^\trans\right). 
\end{align}

When the true density is Gaussian (i.e., $\f = \phi$), we have $\Blf = \Sig^{-1}\Be$ and $\J_\f = \Sig^{-1}$. In this case, the semiparametric SL trace statistic $\Lambda^{SL}_\f$ reduces to the original \textit{Gaussian SL trace statistic}, which is given by
\begin{align} \label{eqn:stat_SL_phi}
\Lambda^{SL}_{\phi} \,:= \, &\trace\,\left(\int_0^1\Be(\s)\dd\big[\Sig^{-1}\Be(\s)\big]^\trans\right)^\trans \nonumber \\
&~~~~~ \times \left(\int_0^1\Be(\s)\Sig^{-1}\Be(\s)^\trans\dd\s\right)^{-1}\left(\int_0^1\Be(\s)\dd\big[\Sig^{-1}\Be(\s)\big]^\trans\right). 
\end{align}

\smallskip
\begin{remark}
Alternatively, one can consider the statistic proposed by \citet{boswijk2015improved}, which corresponds to the case where $\bar\mfC^* = \bar\mfC$.\footnote{After substantial rearrangement, one can achieve $\LR^{\ptrd*}_\f(\bar\mfC;\bar\mfC) = \LR^{\ploc*}_\f(\bar\mfC) + \frac{1}{2}\LR^{d}_\f(\bar\mfC)-\frac{1}{2}\LR^{d}_\f(\mfzero)$, where $\LR^{d}_\f(\mfC) = \CS^d_\f(\mfC)^\trans\QV^{dd}_\f(\mfC)^{-1}\CS^{d}_\f(\mfC)$, with $\CS^{d}_\f(\mfC) := \int_0^1\dcu^{\trans}\dd\big(\Blf(\s)+\Sig^{-1}\We(1)\s - \J_\f\mfC\widetilde{\W}_{\ve}(\s)\s - \Sig^{-1}\mfC\widebar{\W}_{\ve}\s\big)$, $\QV^{dd}_\f(\mfC) := \int_0^1\dcu^\trans\J_\f\dcu\dd\s + \widebar{\mfd}_{\mfC}^\trans\left(\Sig^{-1}-\J_\f\right)\widebar{\mfd}_{\mfC}$, and $\widetilde\W_{\ve}(\s) := \We(\s) - \widebar\W_{\ve}$.} By taking the maximum of $\LR^{\ptrd*}_\f(\bar\mfC;\bar\mfC)$ w.r.t.\ $\bar\mfC$, one obtains the \textit{semiparametric BJN statistic}
\begin{align*} 
\Lambda^{BJN}_\f := \max_{\bar\mfC\in\R^{\p\times\p}}\LR^{\ptrd*}_\f(\bar\mfC;\bar\mfC).
\end{align*}
Similarly, when $f$ is Gaussian, the statistic $\Lambda^{BJN}_\f$ reduces to its Gaussian counterpart introduced in their Section 2.2 of their paper. 
\end{remark}

\subsection{Semiparametrically optimal cointegration tests} \label{subsec:test}
This subsection proposes feasible cointegration tests whose powers can achieve the power bounds developed earlier, establishing that these power bounds are sharp (globally in $\f$) and our tests are semiparametrically optimal (pointwise in $\mfC$).

Our tests are constructed using a nonparametric estimate of the unknown score function $\score_\f$. For this estimation, one could consider the sample splitting technique, which requires fairly minimal conditions on $f$ (see \citet{Bickel1982}, \citet{schick1986asymptotically}, and \citet{drost1997adaptive}). However, to improve the finite-sample performance, especially when the sample size is moderately small, we follow the approach of \cite{schick1987note} which uses the full sample without sample splitting. Unlike other works along this line that require the symmetric density assumption (e.g., \citet{kreiss1987adaptive} for the ARMA model and \citet{ling2003adaptive} for the ARMA-GARCH model), our test statistic construction is based on the framework of \citet{koul1997efficient}, which does not need any additional condition on $\f$.

Our estimate for $\score_\f$ is constructed as follows:
\begin{align} \label{eqn:score_hat}
\hat\score_\f(\ve) = - \frac{\nabla\hat\f(\ve)}{\hat{f}(\ve)+b_\n},
\end{align} 
where $\hat\f$ is the usual kernel density estimator given by
\begin{align} \label{eqn:f_hat}
\hat\f(\ve) = \frac{1}{\n a_\n^\p}\sum_{\ii=2}^{\n}K\left(\frac{\ve-\Delta\vy_\ii}{a_{\n}}\right).
\end{align}
Here $K(\ve) = k(\e_1) \cdots k(\e_\p)$ is the kernel function, and $a_\n$ and $b_\n$ are positive sequences converging to zero. We impose the following mild assumptions.

\smallskip
\begin{assumption} \label{assm:score_hat}
\begin{itemize}
\item[(a)] The marginal kernel function $k(\cdot)$ is bounded, symmetric, continuously differentiable with $\int_{\R}r^2k(r)dr < \infty$ and, for some positive constant C, $|\dot{k}(r)|/k(r) < \infty$ for all $r\in\R$. 
\item[(b)] The sequences $\langle a_\n \rangle$ and $\langle b_\n \rangle$ satisfy $a_\n \to 0$, $b_\n \to 0$, and $\n a_\n^4 b_\n^2 \to \infty$.
\end{itemize}
\end{assumption}
\smallskip

Using the nonparametric estimate $\hat\score_\f$, we construct estimators for $\Blf$ and $\J_\f$ as follows:
\begin{align} \label{eqn:Blfhat}
\Blfhat(\s) := \frac{1}{\sqrt{\n}}\sum_{\ii=2}^{\lfloor\s\n\rfloor}\hat{\score}_{\f}^{(\n)}(\Delta\vy_\ii) - \frac{\lfloor\s\n\rfloor}{\n^{3/2}}\sum_{\ii=2}^{\n}\hat{\score}_{\f}^{(\n)}(\Delta\vy_\ii), ~~  \s\in [0,1],
\end{align}
and
\begin{align} \label{eqn:Ifhat}
\widehat\J_\f := \hat{\score}_\f(\ve_1)\hat{\score}_\f(\ve_1)^{\trans}.
\end{align}
We also define
\begin{align} \label{eqn:Wehat}
\WTe(\s) := \frac{1}{\sqrt{\n}}\sum_{\ii=2}^{\lfloor\s\n\rfloor}\Delta\vy_\ii,
\end{align}
and assume that $\widehat\Sig$ is some consistent estimate of $\Sig$. With these tools in hand, we can define $\widehat\LR^{\ploc*}_\f(\mfC)$ as the feasible version of the likelihood ratio statistic $\LR^{\ploc*}_\f(\mfC)$ by replacing $\We$, $\Blf$, $\Sig$, and $\J_\f$ in the latter with their finite-sample counterparts $\WTe$, $\Blfhat$, $\widehat\Sig$, and $\widehat{\J}_\f$, respectively. In similar way, we define $\widehat\LR^{\ptrd*}_\f(\bar\mfC;\bar\mfC^*)$ as the feasible version of $\LR^{\ptrd}_\f(\bar\mfC;\bar\mfC^*)$. The following theorem summarizes the consistency results for these nonparametric estimators. The proof is presented in Appendix~\ref{I-appendix:mainproofs}.

\smallskip
\begin{theorem} \label{thm:liki_convergence}
\begin{itemize}
\item[(a)] Assuming that $\f\in\F_\p$ and Assumption~\ref{assm:score_hat} holds, let $\Blfhat$ and $\widehat\J_\f$ be defined as in (\ref{eqn:Blfhat})--(\ref{eqn:Ifhat}). Then, as $\n\to\infty$ under $\law_{\mfC,\mfzero;\f}$, we have
\begin{align}
\Blfhat \wto \Blf  {\rm ~~and~~}
\widehat\J_\f \pto \J_\f. 
\end{align}
\item[(b)] Let $\Wehat$ be defined by (\ref{eqn:Wehat}) and let $\widehat\Sig$ be some consistent estimator for $\Sig$ such that $\widehat\Sig\pto\Sig$. Then, under $\law_{\mfC,\lptrd,\lpptb;\ploc,\f}$ and as $\n\to\infty$, 
\begin{align}
\widehat\LR^{\ploc*}_\f(\bar\mfC) \wto \LR^{\ploc*}_\f(\bar\mfC)  {\rm ~~and~~}
\widehat\LR^{\ptrd*}_\f(\bar\mfC;\bar\mfC^*) \wto \LR^{\ptrd*}_\f(\bar\mfC;\bar\mfC^*),
\end{align}
where $\LR^{\ploc*}_\f(\bar\mfC)$ and $\LR^{\ptrd*}_\f(\bar\mfC;\bar\mfC^*)$ are defined respectively in (\ref{eqn:likelihood_MI_mu}) and (\ref{eqn:likelihood_MI_tau}), whose behaviors are characterized by Proposition~\ref{prop:structural_representation}.
\end{itemize}
\end{theorem}
\smallskip

In the proof, a key step is to address the bias, denoted as $\bias$, that arises when estimating $\score_\f$ due to the lack of an $\sqrt{\n}$-unbiased estimator for $\score_\f$ when $\f$ is not further restricted, such as being symmetric (see \cite{Klaassen1987}). However, this bias is canceled out by $\Blfhat$ as shown in the following calculation:
\begin{align*}
&~ \frac{1}{\sqrt{\n}}\sum_{\ii=2}^{\lfloor\s\n\rfloor}\hat{\score}_{\f}^{(\n)}(\Delta\vy_\ii) - \frac{\lfloor\s\n\rfloor}{\n^{3/2}}\sum_{\ii=2}^{\n}\hat{\score}_{\f}^{(\n)}(\Delta\vy_\ii) \\
=&~ \frac{1}{\sqrt{\n}}\sum_{\ii=2}^{\lfloor\s\n\rfloor}(\bias+\check{\score}_{\f}^{(\n)}(\Delta\vy_\ii)) - \frac{\lfloor\s\n\rfloor}{\n^{3/2}}\sum_{\ii=2}^{\n}(\bias+\check{\score}_{\f}^{(\n)}(\Delta\vy_\ii)) \\
=&~ \frac{1}{\sqrt{\n}}\sum_{\ii=2}^{\lfloor\s\n\rfloor}\check{\score}_{\f}^{(\n)}(\Delta\vy_\ii) - \frac{\lfloor\s\n\rfloor}{\n^{3/2}}\sum_{\ii=2}^{\n}\check{\score}_{\f}^{(\n)}(\Delta\vy_\ii),
\end{align*}
where $\check\score_\f^{(\n)}$ denotes the debiased version of $\hat\score_\f^{(\n)}$. Notably, as $\Blf$ eliminates the density perturbation $\lpptb$ in the limit, its finite-sample counterpart $\Blfhat$ eliminates the nonparametric estimation bias in the sequence, effectively addressing the bias issue in the estimation of $\score_\f$.

As a direct consequence of Theorem~\ref{thm:liki_convergence}, the following corollary demonstrates that the power upper bounds $\power^{\ploc*}_{\f,\alpha}(\mfC;\mfC)$ and $\power^{\ptrd*}_{\f,\alpha}(\mfC;\mfC)$ are globally sharp in $\f\in\F_\p$.

\smallskip
\begin{corollary} \label{cor:optimaltests}
Let $\f\in\F_\p$ and assume that Assumption~\ref{assm:score_hat} holds. Define likelihood ratio tests as $\hat\test^{\ploc*}_{\alpha}(\bar\mfC) := \indicator\{\widehat\LR^{\ploc*}_\f(\bar\mfC) > \cv^{\ploc}_\alpha(\bar\mfC)\}$ and $\hat\test^{\ptrd*}_{\alpha}(\bar\mfC) := \indicator\{\widehat\LR^{\ptrd*}_\f(\bar\mfC;\bar\mfC) > \cv^{\ptrd}_\alpha(\bar\mfC)\}$. Then, under $\law_{\mfC,\lptrd,\lpptb;\ploc,\f}$, we have 
\begin{align*}
\lim_{\n\to\infty} \E\big[\hat\test^{\ploc*}_{\alpha}(\bar\mfC)\big] =  \power^{\ploc*}_{\f,\alpha}(\bar\mfC;\mfC) {\rm ~~and~~}
\lim_{\n\to\infty} \E\big[\hat\test^{\ptrd*}_{\alpha}(\bar\mfC)\big] =  \power^{\ptrd*}_{\f,\alpha}(\bar\mfC;\mfC),
\end{align*}
for all $\bar\mfC,\mfC\in\R^{\p\times\p}$.
\end{corollary}
\smallskip

To avoid the necessity of selecting a reference alternative $\bar\mfC$, we adopt the approach used in the aforementioned limiting Johansen test $\Lambda^{Johansen}_\f$ and the limiting SL test $\Lambda_\f^{SL}$. In other words, we construct the feasible versions of these tests by replacing $\We$, $\Blf$, $\Sig$, and $\J_\f$ with their finite-sample counterparts $\WTe$, $\Blfhat$, $\widehat\Sig$, and $\widehat\J_\f$, respectively. These feasible versions are based on the estimate $\hat\f$ and are denoted as $\hat\Lambda^{Johansen}\f$ and $\hat\Lambda\f^{SL}$, respectively.

\smallskip
\begin{remark}
In addition to the kernel estimation method used in this paper, an alternative approach for density estimation is the semi-nonparametric (SNP) method proposed by \cite{gallant1987semi}. For the SNP method employed for the cointegration testing problem, see \cite{boswijk2002semi}.
\end{remark}

\section{Simulations} \label{sec:montecarlo}
In this section, we conduct a Monte Carlo study to validate the asymptotic results using large-sample simulations and evaluate the small-sample performance of our proposed semiparametrically optimal cointegration tests above. Specifically, we compare our tests, which are based on the statistics $\Lambda^{Johansen}_\f$ for the case without time trend, and $\Lambda^{SL}_\f$ for the case with time trend, to their original Gaussian versions, which are based on the statistics $\Lambda^{Johansen}_{\phi}$ and $\Lambda^{SL}_{\phi}$, respectively.

We consider a two-dimension VAR model in error correction form (\ref{eqn:model_1})--(\ref{eqn:model_2}) with i.i.d.\ innovations $\ve_\ii$ of density $f\in\F_2$. Following the asymptotic local power analysis, we set the parameter of interest to be $\mPi = \mfC/\n$ as in (\ref{eqn:ptb_mPi}). We specify the local parameter $\mfC$ using the form
\begin{align*}
\mfC=c\begin{pmatrix}1&1\\0&0\end{pmatrix},
\end{align*}
where $c\in(-\infty,0]$. Thus, $c=0$ corresponds to the null hypothesis of $r=0$, while $c<0$ corresponds to the alternative hypothesis of $r=1$, recalling that $r$ denotes the rank of $\mPi$. We consider $\n=2500$ for the large-sample case and $\n=250$ for the small-sample case. We base all results on $20,000$ independent replications, and set the significance level to be $5\%$ throughout this section.

We estimate the covariance matrix $\Sig$ using the following estimator:
\begin{align*}
\widehat\Sig = \frac{1}{\n-1}\sum_{t=2}^{\n}\left(\Delta\vy_\ii - \overline{\Delta\vy}\right)^\trans\left(\Delta\vy_\ii - \overline{\Delta\vy}\right),
\end{align*}
where $\overline{\Delta\vy} = \sum_{t=2}^{\n}\Delta\vy_\ii/(\n-1)$. To construct the density estimator in (\ref{eqn:f_hat}), which is used to obtain $\score_\f$ in (\ref{eqn:score_hat}) and subsequently $\Blfhat$ and $\widehat\J_\f$ as defined in (\ref{eqn:Blfhat})--(\ref{eqn:Ifhat}), we choose standard Logistic kernels for $k(\cdot)$ and set $b_\n = 0$. Note that the choice of $b_\n = 0$ violates Assumption~\ref{assm:score_hat}-(b), but we made this choice for simplicity and concreteness as the qualitative results appeared to be more sensitive to the choice of $a_\n$ than to the choice of $b_\n$. Following Silverman's rule of thumb (see \cite{silverman1986density}), we set the bandwidth $a_\n$ for each dimension $i = 1,\dots,\p$ to be
\begin{align*}
a_{i,\n} = \left[\frac{4}{\n(\p+2)}\right]^{\frac{2}{\p+4}}\hat\sigma_i^2,
\end{align*}
where $\hat{\sigma}_i^2$ is the $i$-th diagonal element of $\widehat\Sig$.

\begin{figure}[!htb] 
\hspace*{-12mm}
\includegraphics[width=7in]{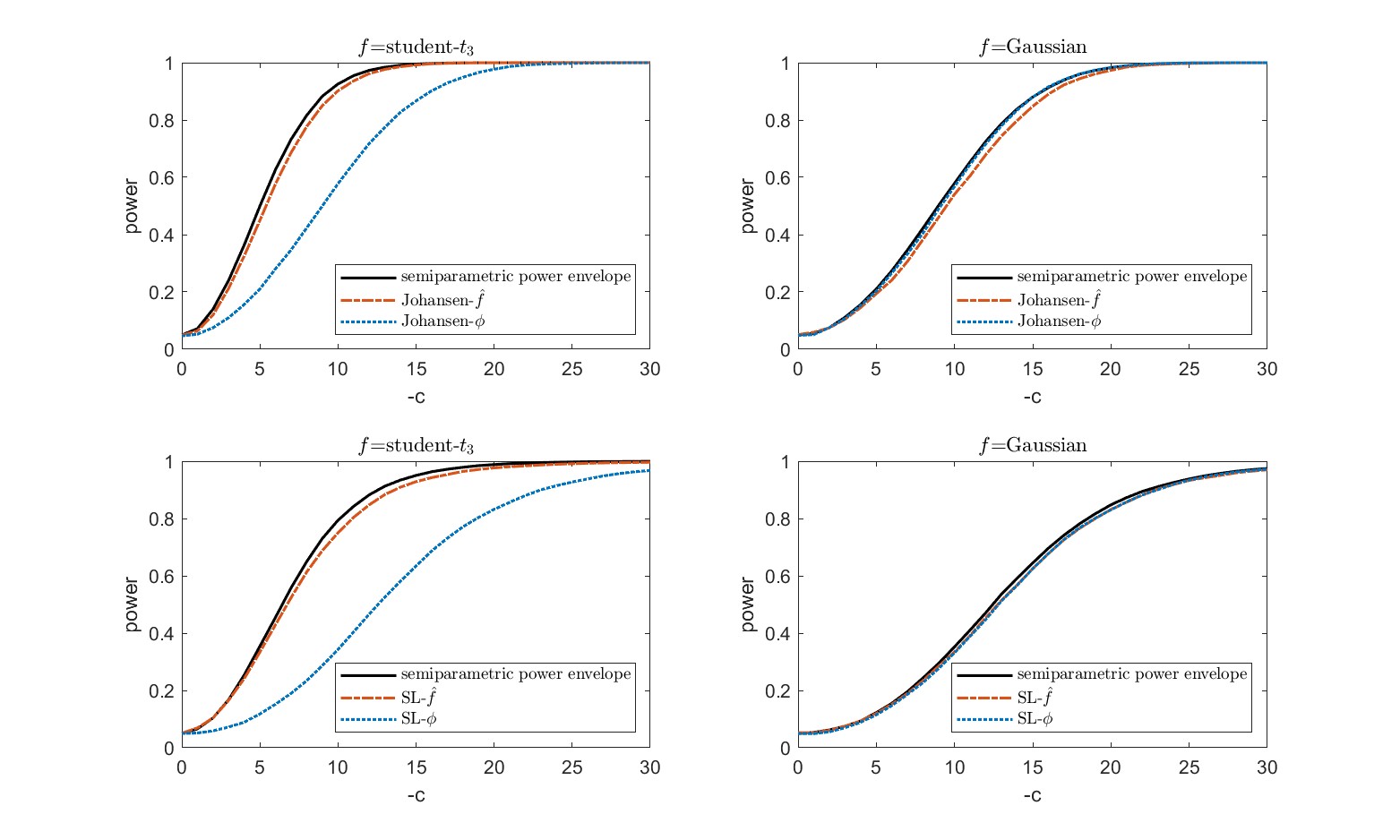}
\caption{Semiparametric power envelope (black solid) and large-sample ($\n = 2500$) rejection rates of the Johansen test (blue dotted) and the $\hat{f}$-based Johansen test (red dash-dot) under true density $\f = student~t_3$ (left panel) and $f = Gaussian$ (right panel).}
\label{figure:plot_T2500}
\end{figure}

\begin{figure}[!htb] 
\hspace*{-12mm}
\includegraphics[width=7in]{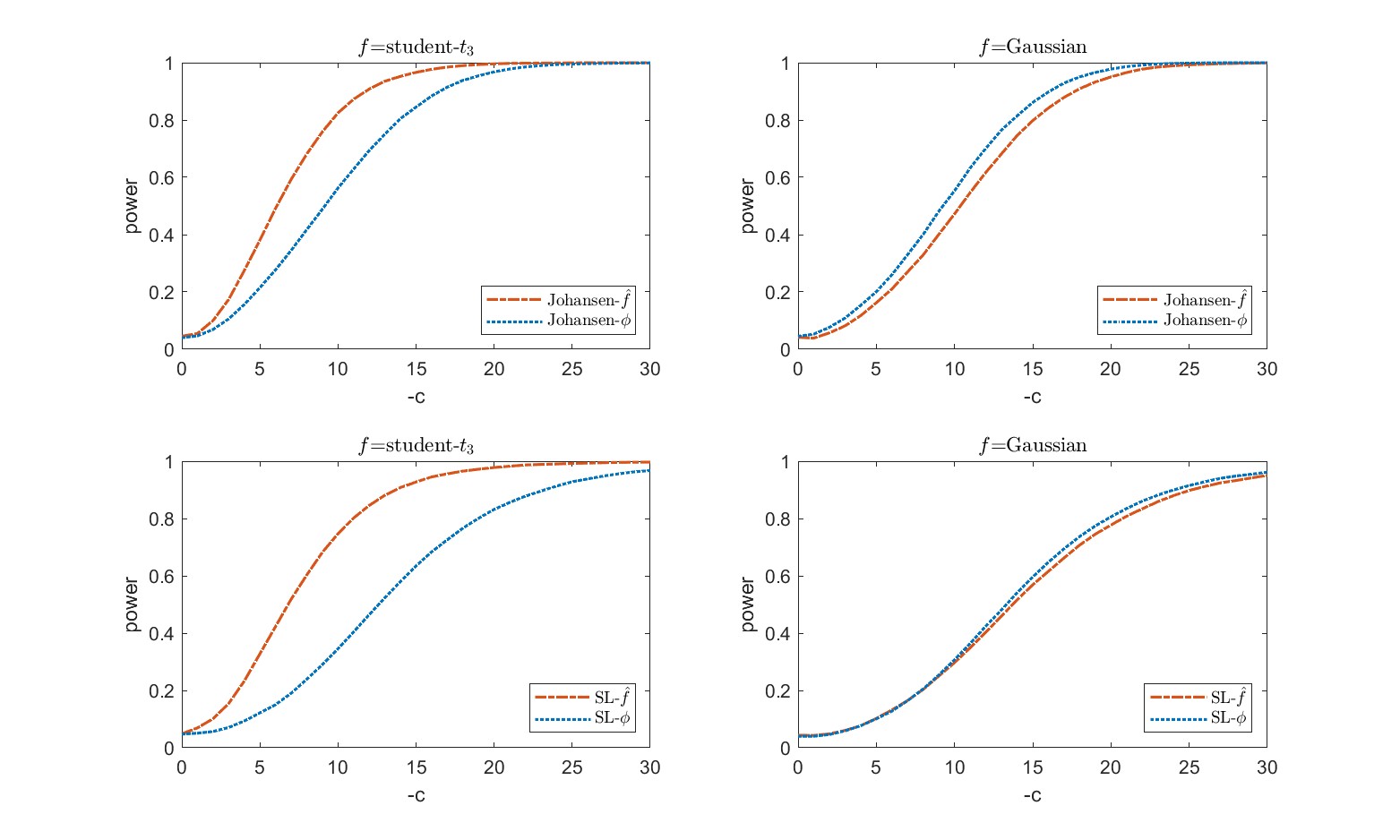}
\caption{Small-sample ($\n = 250$) rejection rates of the Johansen test (blue dotted) and the $\hat{f}$-based Johansen test (red dash-dot) under true density $\f = student~t_3$ (left panel) and $f = Gaussian$ (right panel).}
\label{figure:plot_T250}
\end{figure}

Figure~\ref{figure:plot_T2500} displays the large-sample performances ($\n = 2500$) of the Johansen test (labeled Johansen-$\phi$) and the $\hat{f}$-based Johansen test (labeled Johansen-$\hat{f}$) for the case without a time trend (upper panel), and the SL test (labeled SL-$\phi$) and the $\hat{f}$-based SL test (labeled SL-$\hat{f}$) for the case with a time trend (bottom panel). We consider two different distributions: the Multivariate Student-$t_3$ distribution (left panel) and the Gaussian distribution (right panel). For the case without a time trend and under Gaussian $f$, both Johansen-type tests exhibit similar powers that are close to the Gaussian power envelope. Under the Student-$t_3$ distribution, the $\hat{f}$-based Johansen test outperforms the original Johansen test by a considerable amount of power gain. For instance, at $-c = 10$, the rejection rate of Johansen-$\phi$ is slightly below $60\%$, while that of Johansen-$\hat{f}$ is nearly $90\%$. This demonstrates that our semiparametric efficient Johansen test can effectively utilize the information contained in the innovation distribution, which is particularly advantageous when the true density deviates significantly from Gaussian, a limitation of the original version.

Furthermore, it is worth noting that a small discrepancy may be observed between the power of the Johansen-$\hat{f}$ test and the semiparametric power envelope. This slight reduction in power can be attributed to the finite-sample nature of the data, even with a relatively large sample size, and the nonparametric estimation of the density $f$. Nevertheless, this gap tends to decrease as the sample size $T$ increases towards infinity, and the power of the Johansen-$\hat{f}$ test converges to the semiparametric power envelope. These results are also applicable to the time trend case for the corresponding SL-$\phi$ and SL-$\hat{f}$ tests, which provides numerical evidence that our proposed $\hat{f}$-based tests are semiparametrically optimal.

Figure~\ref{figure:plot_T250} presents the small-sample performances with a sample size of $n=250$, which corresponds to the scenarios depicted in Figure~\ref{figure:plot_T2500}. These results confirm that all four tests under evaluation, namely Johansen-$\phi$ and Johansen-$\hat{f}$ for the no-time-trend case, and SL-$\phi$ and SL-$\hat{f}$ for the time-trend case, maintain satisfactory size control even with a relatively small sample size. Furthermore, we observe that the power properties observed in the large-sample case are also applicable in the small-sample case. Notably, the $\hat{f}$-based tests provide significant power improvement over their Gaussian counterparts when $f$ follows a Student-$t_3$ distribution, even with a small sample size. However, when $f$ is Gaussian, the $\hat{f}$-based tests may exhibit slightly lower power compared to their Gaussian counterparts, mainly due to the need for nonparametric density estimation.

\begin{figure}[!htb] 
\hspace*{-12mm}
\includegraphics[width=7in]{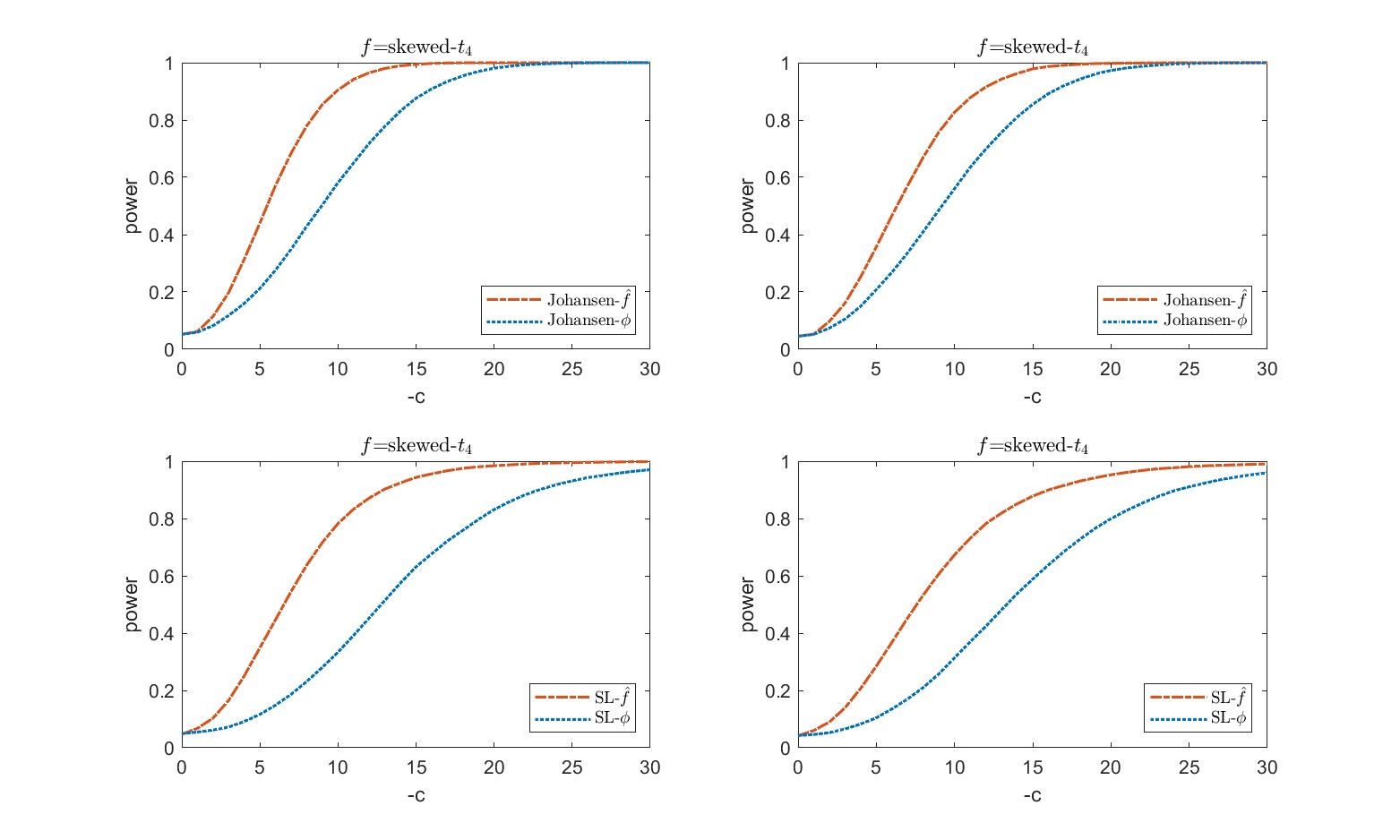}
\caption{Large-sample ($\n = 2500$, left panel) and small-sample ($\n = 250$, right panel) rejection rates of the Johansen test (blue dotted) and the $\hat{f}$-based Johansen test (red dash-dot) under true density $\f = skewed~t_4$.}
\label{figure:plot_skewt4}
\end{figure}

We extend our investigation to the scenarios under asymmetric distributions, as illustrated in Figure~\ref{figure:plot_skewt4}. Specifically, we examine the performance of the four tests (Johansen-$\phi$, Johansen-$\hat{f}$, SL-$\phi$, and SL-$\hat{f}$) under the Skewed-$t_4$ distribution in both large ($T=2500$, left panel) and small ($T=250$, right panel) samples for both the no-time-trend case (upper panel) and the time-trend case (bottom panel). Our results confirm the previously reported findings, including satisfactory size control and the remarkable power improvement of the $\hat{f}$-based tests over their Gaussian counterparts, are applicable to the asymmetric distribution case as well. Moreover, additional simulations unreported here indicate that the $\hat{f}$-based tests gain more and more power as the Skewed-$t_4$ distribution becomes more and more skewed (i.e., further away from Gaussian).



\section{Discussions on extensions} \label{sec:extensions}
The preceding sections have focused on a simple case for semiparametric efficiency, which assumed no serial correlation in the error term and considered only the null hypothesis of no cointegrating relationship (i.e., $H_0: \, r = 0$). To broaden the applicability of our tests, we briefly discuss whether our results remain valid when we relax the assumption of no serial correlation and consider a more general reduced rank hypothesis.\footnote{In the Gaussian case, \citet{boswijk2015improved} have addressed the results of relaxing these two assumptions in their Sections 2.3 and 2.4.}

We consider a generalized model where the observations $\vy_\ii = (\y_{1,\ii},\dots,\y_{\p,\ii})^{\trans}$ are generated as follows:
\begin{align} 
\vy_\ii =&~ \ploc + \ptrd\ii + \vx_\ii,  \\
\Delta\vx_\ii =&~ \mPi\vx_{\ii-1} + \sum_{j=1}^{k-1}\mGamma_j\Delta\vx_{t-j} + \ve_\ii, 
\end{align}
where, in addition, the parameter $\mGamma = \{\mGamma_1,\dots,\mGamma_{k-1}\}\in\R^{p\times(k-1)p}$ of (known and finite) order $p$ governs the lag terms. We are interested in testing the hypothesis 
\begin{align*}
H_0:r=r_0 {~~\rm against~~} H_1:r>r_0, 
\end{align*}
where $r_0\in\N_0$ and $r_0 < \p$.

Our discussion is based on the results in HvdAW, specifically their `complete' version of the limit experiment for cointegration provided in their online supplementary appendix (Proposition A.2). Following HvdAW, we consider the following localized parameterizations:
\begin{itemize}
\item For $\mGamma$, we use
\begin{align}
\mGamma = \mGamma^{(\n)}_{\mfG} = \mGamma_0 + \frac{\mfG}{\sqrt{\n}},
\end{align}
where the local parameter $\mfG = \{\mfG_1,\dots,\mfG_{k-1}\} \in \R^{\p\times(k-1)\p}$
\item For $\mPi$, building upon the factorization $\mPi = \malpha_0\mbeta_0$, where $\malpha_0$ and $\mbeta_0$ are full-rank $\p\times r_0$ matrices, we consider
\begin{align*}
\mPi = \mPi^{(\n)}_{\mfA,\mfB,\mfC} = \malpha\mbeta^\trans+\frac{\malpha_\perp\mfC\malpha_\perp^\trans}{\n},
\end{align*}
and
\begin{align}
\malpha = \malpha^{(\n)}_{\mfA} = \malpha_0 + \frac{\mfA}{\sqrt{\n}} {\rm ~~~and~~~} \mbeta = \mbeta^{(\n)}_{\mfB} = \mbeta_0 + \frac{\mbeta_{\perp}\mfB^\trans}{\n}
\end{align}
where $\malpha_\perp$ and $\mbeta_\perp$ are chosen $\p\times(\p-r_0)$ matrices of rank $\p-r_0$ satisfying $\malpha^\trans\malpha_\perp = \mfzero_{r_0\times(\p-r_0)}$ and $\mbeta^\trans\mbeta_\perp = \mfzero_{r_0\times(\p-r_0)}$, and $\mfC\in\R^{p-r_0\times p-r_0}$. 
\end{itemize}
With these local alternatives, our inference problem becomes testing the null hypothesis of $\mfC=\mfzero$ (since $r=r_0$ if and only if $\mfC=\mfzero$), while treating $\mfG$, $\mfA$ and $\mfB$ as nuisance parameters.

We incorporate our generalized model into the framework of \citet[Proposition A.2]{hallin2016semiparametric}. Recalling above that the key difference between our model and theirs is the specification of the trend term, with our model described by equations (\ref{eqn:model_1})--(\ref{eqn:model_2}), while theirs is described by equation (\ref{eqn:model_h}). Our case corresponds to theirs when their parameter $\mu = 0$ (and our additional time trend term $\ploc + \ptrd\ii$ can be handled in a similar manner as described in Section~\ref{subsec:eliminate_lptrd}). As a result, all local perturbations associated with $\mu$, specifically their parameters $b$, $d$, and $\mu$ itself, are absent in our model. While HvdAW focus on the parameter of interest $d$, which enjoys a ``super-consistency'' rate $T^{-3/2}$ and the traditional LAN result, our paper focuses on the parameter $\mfC$ (corresponding to their $D$), which leads to the LABF result and was left uninvestigated in their work.

Another difference between HvdAW and our present paper is that the former assumes $\f$ is elliptical, while we do not make this assumption. The elliptical density assumption in HvdAW is specifically used for their test construction based on the pseudo-Mahalanobis distance-based rank statistic. However, it is not necessary for the limit experiment approach to proceed. Therefore, it is reasonable to conjecture that their limit experiment has the same structure without this distribution shape restriction. In the following, we will translate their results by replacing their $W_{\epsilon}$ and $W_{\phi}$ with our $\We$ and $\Wlf$, respectively, which are both $\p$-dimensional Brownian motions that play the role of the limits of the partial-sum processes of the innovations and the score functions.

To save space, we will not present a full exposition of the limit experiment, nor provide a rigorous proof for it (which could be done similarly to Proposition~\ref{prop:limitexperiment_LABF}). Instead, we only list the limiting central sequences with respect to $({\rm vec}\,\mfC)^\trans$ $({\rm vec}\,\mfG)^\trans$, $({\rm vec}\,\mfA)^\trans$ and $({\rm vec}\,\mfB)^\trans$, respectively, as follows (borrowed from \citet[Proposition A.2]{hallin2016semiparametric}):
\begin{align*}
\CS_\mfC &= \int_0^1\big(\mbeta_\perp^\trans\mfD\We(\s)\otimes\id_{\p-r_0}\big)\dd\big(\malpha_\perp^\trans\Wlf\big)(\s), \\
\CS_\mfA &= \big(\mbeta^\trans\otimes\id_{\p}\big)\W_{1}(1), \\
\CS_\mfB &= \int_0^1\big(\mbeta_\perp^\trans\mfD\We(\s)\otimes\id_{r_0}\big)\dd\big(\malpha^\trans\Wlf\big)(\s), \\
\CS_\mfG &= \W_{2}(1), 
\end{align*}
where $\mfD = \mfD_{\mGamma,\mPi} := \mbeta_\perp\left(\malpha_\perp^\trans(\id_\p - \sum_{j=1}^{k-1}\mGamma_j)\mbeta_\perp\right)^{-1}\malpha_\perp^\trans$, and $\W_1$ and $\W_2$ are Brownian motions defined in the same space of $(\We,\Wb,\Wlf)$.

According to the covariance analysis of HvdAW (as discussed above their Lemma A.1), it can be shown that $\W_2$ (referred to as $W_{\Delta X\otimes\phi}$ in HvdAW) is uncorrelated with $\We$ and $\Wlf$ when their $\mu$ is zero.  This can be explained by observing that the score associated with $\W_2$ is constructed by multiplying $\score_\f(\Delta\vx_\ii)$ with the lagged increments, $\Delta\vx_{\ii-1},\dots,\Delta\vx_{\ii-k}$, which is independent of any function of $\Delta\vx_\ii$, and has mean zero under the null hypothesis.\footnote{It is worth noting that this property is also shared by our univariate counterpart, \citet{zhou2019semiparametrically}, where $\W_2$ here corresponds to the $W_\Gamma$ there, and the latter is uncorrelated with the univariate counterparts of $\We$ and $\Wlf$ as described in (3.8) of their paper.} As a result, $\CS_\mfC$ is independent of $\CS_\mfG$, which implies that the inference for $\mfC$ is adaptive with respect to $\mfG$. In other words, the parameters $\mGamma$ that control the serial correlation in the error term can be treated ``as if'' they are known, and one can simply replace them with their consistent estimates. Unreported simulation results with ARMA errors can confirm our conjecture (these results are available upon request).

The adaptivity result also applies to $\malpha$ and $\mbeta$, as it can be easily shown that $\CS_\mfC$ is independent of both $\CS_\mfA$ (due to $\mbeta^\trans\mbeta_\perp = \mfzero$) and $\CS_\mfB$ (due to of $\malpha^\trans\malpha_\perp = \mfzero$). Hence, by the same reasoning, one can simply substitute these parameters with their consistent estimates to achieve feasible inference for $\mfC$ without sacrificing asymptotic efficiency. In summary, for this general reduced null hypothesis case, one can follow these steps: (i) estimate $\alpha$ and $\beta$ based on the original data $\vy_\ii$, obtain an estimator $\hat{\alpha}\perp$ for $\alpha\perp$ using the orthogonality of $\alpha$ and $\alpha_\perp$ based on the estimates from step (i), and (iii) apply the our proposed semiparametric optimal tests to the transformed data $\hat{\alpha}_\perp\vy_\ii$. This procedure is similar to the one outlined in \citet[Section 2.3]{boswijk2015improved} except for step (iii), except that we use our proposed tests in step (iii).

\bibliographystyle{asa}
\bibliography{references}

\end{document}